\documentstyle[12pt,epsfig,multirow]{article}

\newcommand{\be}{\begin{eqnarray}}
\newcommand{\ee}{\end{eqnarray}}

\def\nue{{\nu_e}}
\def\anue{{\bar\nu_e}}

\newcommand{\pre}{\phi^{r}_{\nu_{e}}}
\newcommand{\prae}{\phi^{r}_{\bar\nu_{e}}}

\def\ltap{\ \raisebox{-.4ex}{\rlap{$\sim$}} \raisebox{.4ex}{$<$}\ }
\def\gtap{\ \raisebox{-.4ex}{\rlap{$\sim$}} \raisebox{.4ex}{$>$}\ }

\begin{document}

\thispagestyle{empty}
\begin{flushright}

\end{flushright}
\bigskip

\begin{center}
{\Large \bf Collective Flavor Oscillations Of Supernova Neutrinos \\and r-Process Nucleosynthesis} 

\vspace{.5in}

{\bf {Sovan Chakraborty$^{\star, a}$, 
Sandhya Choubey$^{\dagger, b}$, Srubabati Goswami$^{\sharp, c}$,
Kamales Kar$^{\star, d}$}} 
\vskip .5cm
$^\star${\normalsize \it Saha Institute of Nuclear Physics,}\\
{\normalsize \it 1/AF Bidhannagar, Kolkata 700064, India}\\
\vskip 0.4cm
$^\dagger${\normalsize \it Harish-Chandra Research Institute,} \\
{\normalsize \it Chhatnag Road, Jhunsi, Allahabad  211019, India}\\
\vskip 0.4cm
$^\sharp${\normalsize \it Physical Research Laboratory, Navrangpura,}\\
{\normalsize \it Ahmedabad 380009, India} 
\vskip 1cm

{\bf ABSTRACT}
\end{center}

Neutrino-neutrino interactions inside core-collapse supernovae may give rise
to collective flavor oscillations resulting in swap between flavors. 
These oscillations depend on the initial energy spectra, 
and relative fluxes or relative luminosities of the neutrinos.
It has been observed that 
departure from energy equipartition among different flavors 
can give rise to   
one or more sharp spectral swap over energy, termed as splits. 
We study the occurrence of splits  
in the neutrino and antineutrino
spectra, varying the initial relative fluxes for different models of 
initial energy spectrum, in both normal and inverted hierarchy. These initial
relative flux variations give rise to several possible split patterns whereas
variation over different models of energy spectra give similar results. 
We explore the effect of these 
spectral splits  on the electron fraction, $Y_e$,
that governs r-process nucleosynthesis inside supernovae. 
Since spectral splits modify the electron neutrino and 
antineutrino spectra in the region where r-process is 
postulated to happen, and since the pattern of spectral 
splits depends on the initial conditions of the spectra 
and the neutrino mass hierarchy, we show that 
the condition $Y_e < 0.5$ required 
for successful
r-process nucleosynthesis will lead to constraints on the 
initial spectral conditions, for a given neutrino mass hierarchy. 

\noindent

\vfill

\noindent $^a$ email: sovan.chakraborty@saha.ac.in

\noindent $^b$ email: sandhya@hri.res.in

\noindent $^c$ email: sruba@prl.res.in

\noindent $^d$ email: kamales.kar@saha.ac.in

\newpage

\section{Introduction}

Neutrinos from a core-collapse supernova can play an important role
in probing both neutrino properties as well as throwing light on
the supernova mechanism \cite{Raff, Dighe:2008dq}.
Neutrinos emitted during the explosion of
core-collapse supernovae (SN) pass through very large density variation of
matter and can undergo  MSW  resonant flavor 
conversion \cite{msw}  which can give useful information on 
neutrino mass hierarchy and the third leptonic mixing angle $\theta_{13}$. 
They are also influenced by
the shock wave formed in the SN core and thus carry information about the
explosion mechanism \cite{fogli-lisi-mirizzi-montanino-0304056, dasgupta-dighe-0510219, choubey-harries-ross-0605255, fogli-lisi-mirizzi-montanino-0603033,schirito-Fuller-0205390,Tomas2004}.

Recently it was realized that
a crucial feature in the study of 
SN neutrinos
comes from the collective 
neutrino-neutrino interaction at very high densities of
the core and this may change the emitted flux of different flavors
substantially  \cite{Pantaleone:1992eq}-\cite{Fogli:2009rd}.
The phenomenology of supernova neutrinos with collective effect
including Earth matter effect on SN neutrinos \cite{Dasgupta:2008my},
prompt SN \cite{Duan:2007sh, Dasgupta:2008cd}, 
diffuse neutrino background from SN \cite{Chakraborty:2008zp}, 
failed SN \cite{Lunardini:2009ya} as well as
CP violation in SN neutrinos \cite{Gava:2008rp}
has been studied.
For a recent review we refer to \cite{Duan:2009cd}.

It has been shown that effectively the collective evolution of a three-flavor 
($\nu_{e}, \nu_{\mu},\nu_{\tau} $) system can be treated like a two flavor 
($\nu_{e}, \nu_{x} $) scenerio, where  $\nu_{x}$ can be $\nu_{\mu}$ or 
$\nu_{\tau}$ or a linear combination of $\nu_{\mu}$ and $\nu_{\tau}$ \cite{Dasgupta:2007ws, Fogli:2008fj}.
The flavor evolution for this system is driven by the 
effective mass squared difference  
$\Delta m^2$ and the mixing angle $\theta_{13}$. 
This two flavor scenerio which has
been extensively studied \cite{Duan:2006an, Raffelt:2007cb, Raffelt:2007xt, Fogli:2007bk}  shows that for 
inverted hierarchy (IH, $\Delta m^2 <0$),  
above a critical energy (split energy $E_{c}$),  the 
spectrum in both the electron neutrino ($\nu_{e}$) and antineutrino 
($\bar\nu_{e}$) sectors end up with a complete exchange or swap with 
$\nu_{x}$ and  $\bar\nu_{x}$ respectively, this is referred as ``spectral swap''. 
Recently the studies in \cite{Dasgupta:2009mg, Fogli:2009rd} 
analyzed the role of equipartition in energy and variation of luminosity and
showed the interesting possibility of multiple splits in the supernova neutrino spectrum for IH. 
Single spectral split for Normal Hierarchy (NH, $\Delta m^2 > 0$) 
was also reported for certain values of luminosities. 

In this paper our goal is to study  (a) the impact of spectral splits on the electron fraction ($Y_e$) which is 
a diagnostic of successful r-process nucleosynthesis in supernova and  
(b) the inverse problem which is to see
if it is possible to put any constraint on  the luminosities by
demanding the neutron rich condition on $Y_e$ which is required for 
synthesis of heavy nuclei.  

Though the site for r-process nucleosynthesis is not known definitely,
supernovae are considered to be excellent candidates for it. One of the
criteria for the rapid nucleosynthesis to take place is that it has to
be in a neutron-rich region. With the two competing beta processes
$n+\nu_e \rightarrow p+e^{-}$ and $p+\bar\nu_e \rightarrow n+e^{+}$
occurring in the hot bubble and neutrino driven wind region, the minimal
condition is that the electron fraction, $Y_e <0.5$. A more realistic  
constraint may be $Y_e <0.45$.

Effect of MSW neutrino oscillations 
on r-process nucleosynthesis were considered in \cite{Qian:1993dg}, 
and subsequently in a large number of papers. Since 
r-process is expected to take place in the neutrino driven wind 
deep inside the supernova, the matter density in these regions 
obviously are high. Therefore, in order to have MSW effect 
one requires mass squared difference $\Delta m^2 \sim 1$ eV$^2$ 
or higher. Such large values of  $\Delta m^2$ are possible only 
if one allows for sterile neutrinos. 
For only active neutrinos, the mass squared differences are 
of the order of $10^{-3}$ eV$^2$ (atmospheric) and $10^{-5}$ eV$^2$
(solar) only, and the MSW resonance for these are reached at 
distances far beyond the r-process region. 
Therefore with only active neutrinos, one expects 
no effect of MSW oscillations 
on r-process nucleosynthesis. 
However, one will have collective effects driven by 
these active neutrino mass squared differences and these 
collective flavor oscillations 
happen very close to the neutrinosphere -- 
close enough to impact r-process nucleosynthesis. 
The effect of the 
collective oscillations on the possibility of getting n-rich region in the hot 
bubble was studied in \cite{Pastor:2002we}. 
The problem of r-process in the
neutrino-wind region with different evolution scenarios
was looked into in \cite{Balantekin:2004ug}. 
With an improved
understanding of the spectral splits in the collective oscillations 
\cite{Dasgupta:2009mg, Fogli:2009rd}, we re-examine the problem of SN
r-process  using the two flavor basis as mentioned above. 
In particular, recent papers have shown that (multiple)  
spectral splits could happen in either or both the neutrino and 
antineutrino channels, depending on the initial spectral 
conditions. Since spectral splits can change the neutrino and 
antineutrino spectra and hence the value of $Y_e$, 
and since these spectral split patterns depend on the initial 
conditions of the unoscillated spectra, 
r-process 
nucleosynthesis can in principle be used to put constraints 
on the initial spectral conditions. We perform this 
exercise as a proof of principle, in a simplified framework, 
in order to illustrate that such constraints indeed exist. 
We consider different models for the initial neutrino spectrum 
with different hierarchies among average energies and departures from 
energy equipartition among flavors. We demonstrate  
the occurrence  of spectral splits due to these variations  
on the probabilities and the fluxes for both hierarchies.  
We use the constraint that the environment should be neutrino rich for 
successful r-process and 
delineate the  resulting constraints 
that are obtained on the fractional luminosities.
  
The paper is organized as follows. In Section 2, we outline 
the neutrino-neutrino interaction in the context of the supernova problem.
First the two flavor evolution equations are presented. 
Next we vary the fractional luminosities for different initial energy 
spectrum model and study the 
the effect of  collective oscillations, especially the effect of 
the spectral splits  on the
emitted neutrino flux. In Section 3, 
we discuss how  the evolution of the electron fraction $Y_e$ is affected 
by collective oscillations.   
The value of $Y_e$ 
determines the possibility of having
r-process nucleosynthesis in the supernova environment. 
We impose 
the constraint that the environment should be neutron rich 
and study the initial relative fluxes or relative luminosities
which are allowed under this constraint.
Finally Section 4 makes some
concluding remarks. 

\section{Two flavor neutrino-neutrino interaction \\and Supernova}

Close to the neutrinosphere, due to the large neutrino density, neutrinos form 
a background to themselves. This neutrino-neutrino interaction effect is 
nonlinear and can give rise to collective flavor transition of neutrinos and 
antineutrinos. In this section we discuss the flavor evolution equations,
different models of supernova neutrino spectrum and the impact of the 
collective effects on the probabilities and the fluxes for both NH and IH. 

\subsection{Two flavor evolution equations of SN Neutrinos }

Due to the large neutrino density inside the neutrinosphere, the
neutrino-neutrino interactions lead to  
coherent oscillations of 
neutrinos of different energies 
with some average frequency, giving rise to synchronized 
oscillations. However, there is no effective flavor conversion 
due to these synchronized 
oscillations as the effective mixing angle is highly suppressed due to the 
large MSW potential in the region close to the neutrinosphere. 
With the neutrino density decreasing outward, bipolar 
oscillations begin to take place. These oscillations can lead to complete or 
partial swapping (spectral split) of the $\anue$ ($\nu_e$) and 
$\bar{\nu}_x$ ($\nu_x$) spectra depending on their 
initial luminosities and 
average energies. 
Finally, after a few hundred kilometers, the neutrino-neutrino
interactions become negligible and it is the MSW transitions 
which dominate. 

As shown in \cite{Dasgupta:2007ws, Fogli:2008fj} the collective effect due to 
neutrino-neutrino interaction effectively involves only two flavors of 
neutrinos ($\nu_{e},\nu_{x}$), 
while the other flavor ($\nu_{y}$) does not evolve under
this collective potential. Here $\nu_{x}$ is a linear combination of 
$\nu_{\mu}, \nu_{\tau}$ and $\nu_{y}$ is the orthogonal combination to 
$\nu_{x}$. The only way $\nu_{y}$ can effect the final neutrino spectrum is by 
MSW transition which happens at a larger radius of about $10^{4-5}$ km, well  
beyond the collective region which is within a few hundred km from 
the center of the exploding star. Therefore, the
effective evolution of the neutrinos 
is very well described by the two flavor formalism.
\footnote{
{
The MSW $\mu$-$\tau$ resonance can occur inside the supernova under certain
  conditions \cite{Kneller:2009}. Unlike the standard MSW resonances involving the
  electron flavor, the $\mu$-$\tau$ resonance can occur in the inner
  supernova layers. The effect of $\mu$-$\tau$ neutrino refraction and
  collective three-flavor transformations in supernovae has been
  considered in \cite{Andreu:2007}.
 We have ignored these effects in this work.

}}

The evolution equations in the two-family Bloch vector notation for the 
polarization vectors of the neutrino 
(\textbf{P}) and antineutrino ($ \textbf{P}^{'} $) sector are,
\be
 \textbf{\.{P}} = \textbf{P}\times(\omega \textbf{B}-\lambda\textbf{\^z}-\mu\textbf{D})~ ,
\ee
\be
 \textbf{\.{P}}^{'} = \textbf{P}^{'}\times(-\omega \textbf{B}-\lambda\textbf{\^z}-\mu\textbf{D}) ~,
\ee
where the terms involving $\omega$, $\lambda$ and $\mu$ are the ones having
the vacuum, matter and neutrino-neutrino interaction effects and the 
frequencies are represented by
\be
\textbf{B}=(-\sin 2\theta, 0,\cos 2\theta)^{T}~~~,~~~\omega = \frac{\Delta{m}^{2}}{2E} ~,
\ee
\be
\textbf{\^{z}} = (0,0,1)^{T}~~~,~~~ \lambda = \sqrt{2}G_{F}N_{e} ~,
\ee 
\be
\textbf{D} =\frac{1}{(N_{\nu_e}+N_{\nu_x}+N_{\bar\nu_e}+N_{\bar\nu_x})}\,{\int}\,dE (n\textbf{P}-\bar{n}\textbf{P}^{'}) ~~~,~~~ \mu = \sqrt{2}G_{F}(N_{\nu_e}+N_{\nu_x}+N_{\bar\nu_e}+N_{\bar\nu_x}) ~,
\ee
respectively.\\
$\textbf{D}$ can be defined in terms of the global polarization vectors $\textbf{J}$ and $\bar \textbf{J} $ i,e ~~~~$ \textbf{D}=\textbf{J} -\bar \textbf{J}~
$.\\
Where\\ 
 $~~~~~~~~~~~~~~ ~\textbf{J}=\frac{1}{(N_{\nu_e}+N_{\nu_x}+N_{\bar\nu_e}+N_{\bar\nu_x})}\,{\int}\,dE~ n\textbf{P} ~$
~~,~~
$ \bar\textbf{J}=\frac{1}{(N_{\nu_e}+N_{\nu_x}+N_{\bar\nu_e}+N_{\bar\nu_x})}\,{\int}\,dE~ \bar{n}\textbf{P}^{'}
$.\\

 As usual, 
$\theta$ and $\Delta m^{2}$ are the mixing angle 
and mass squared difference respectively. 
In what follows
$\theta_{eff}$ is taken as $10^{-5}$ and $\vert \Delta {m}^{2} \vert$ = $\vert \Delta m^{2}_{31} \vert$ = $\vert m^{2}_{3}-m^{2}_{1} \vert$ =$3\times10^{-3}$ eV$^2$. 
$N_{\alpha}$'s represent the total effective number density of the $\alpha$th species.
\be
N_{\alpha}=\,{\int}\,dE ~n_{\alpha} ~,
\ee
where, 
\be
 n = n_{\nu_{e}} + n_{\nu_{x}} ~~~,~~~ \bar{n} = n_{\bar\nu_e} + n_{\bar\nu_x}~,
\ee
$n_{\alpha} $'s are the effective number density per unit
energy for the $\alpha$'th species of neutrino and can be expressed as  
\cite{Duan:2006an}
\be
n_{\alpha}(r,E) = \frac{D(r)}{2\pi R_{\alpha}^{2}}\frac{L_{\alpha}}{\langle{E_{\alpha}}\rangle}\Psi_{\alpha}(E)~,
\ee
where $L_{\alpha}$ and $\langle{E_{\alpha}}\rangle$ are the 
luminosity and average energy for the $\alpha$th (anti)neutrino species,  
$R_{\alpha}$ is the neutrinosphere 
radius. 
The initial flux of the $\alpha$th species at the neutrinosphere is given by $\frac{L_{\alpha}}{\langle{E_{\alpha}}\rangle}$ whereas the initial energy distribution is represented by $\Psi_{\alpha}(E)$.
$D(r)$ denotes 
the geometrical function in the ``single angle approximation'' 
\cite{Duan:2006an} can be expressed as, 
\footnote{
{It is well known that the current-current nature of the weak interaction 
introduces an angular factor in the Hamiltonian because of which 
the neutrinos from 
the SN core traveling along different trajectories  
encounter different neutrino-neutrino interaction potential. 
These `multi angle' effects may give rise to kinematical decoherence
\cite{Raffelt:2007yz} which in turn can wash out the collective features 
described above. But for spherically symmetric cases ``single angle'' 
approximation i,e neutrino-neutrino interactions averaged along a single trajectory seems to be a fine approximation as the `multi angle' decoherence in this case is rather weak against the collective features \cite{EstebanPretel:2007ec}. 
We are aware that the extent of kinetic decoherence is initial flux condition dependent and hence some of the initial fluxes that we consider might lead to kinetic decoherence.
In \cite{Fogli:2009rd} the different possible split patterns with varying
neutrino luminosities were studied using the ``single angle approximation''. 
In this work 
we study the multiple split patterns for different models of 
initial supernova neutrino spectrum  in the same spirit as \cite{Fogli:2009rd}.
Multi angle study of these possible spectral patterns for different initial
supernova neutrino spectrum might give interesting results like the effect of decoherence, but this
is beyond the scope of our work.
}}
\be
D(r)= \frac{1}{2}\left( 1-\sqrt{1-\left( \frac{R_{\alpha}}{r}\right) ^{2}}\right) ^{2} ~.
\ee

The matter effect is removed from the evolution equations as the equations are
considered in a frame rotating with angular velocity -$\lambda\textbf{z}$
\cite{Duan:2005cp}.
\footnote{Note that though the matter potential is mostly rotated away, 
it may affect the evolution by delaying the collective effect
\cite{Hannestad:2006nj} or by some early decoherence 
\cite{EstebanPretel:2008ni} or even modifying very low energy 
(order of 0.1 MeV) split features \cite{Fogli:2007bk,Fogli:2008pt,Dighe:2008dq}.These early effects have very little impact on the over all split patterns 
at the end of the collective region (400 Km) and the low energy split features 
below 1 MeV are negligible compared to the total spectra. 
Moreover the matter potential can be accounted for by choosing a matter 
suppressed hence small effective mixing angle, as we have chosen a  
$\theta_{eff}$ = $10^{-5}$ \cite{EstebanPretel:2007ec}.    
So in the subsequent discussions we neglect the above mentioned roles of the
matter term and work with a very small $\theta_{eff}$ to compensate the 
matter term. We have explicitly checked that the inclusion of the matter term 
does not change our results. } 
In such a frame all the physical observable remain the 
same.
Thus the evolution equations are 
\be
\textbf{\.{P}} = \textbf{P}\times(\omega \textbf{B}-\mu\textbf{D})~,
\ee
\be
 \textbf{\.{P}}^{'} = \textbf{P}^{'}\times(-\omega \textbf{B}-\mu\textbf{D})~.
\ee
These are nonlinear coupled equations (due to the 2nd term containing $\textbf{D}$) and have to be solved numerically.
It is evident from the evolution Eqs. (10) and (11) that there are two 
relevant frequencies, the usual vacuum frequency ($\omega$) and the
neutrino-neutrino interaction strength parameter ($\mu$).
For our chosen $\Delta m^{2}$ the
vacuum frequency is
\be
\omega = \frac{\Delta{m}^{2}}{2E} = \frac{30}{(4E/{MeV})}~~ km^{-1}~.
\ee
The usual SN neutrino energy considered is in the range 
of 0 to 50 MeV, as the SN neutrino flux beyond 50 MeV 
is very small. Hence we use neutrino energy upto 50 MeV for our calculation.

The other frequency ($\mu$) representing neutrino-neutrino interaction is 
nontrivial, and is given by
\be
\mu = \sqrt{2}G_{F}(N_{\nu_e}+N_{\nu_x}+N_{\bar\nu_e}+N_{\bar\nu_x})~.
\ee
The Eqs. (6) to (9) imply that contribution of the $\alpha$-th species to 
$\mu$ is dependent on radial distance ($r$), 
neutrinosphere radius ($R_{\alpha}$),
initial flux ($\frac{L_{\alpha}}{\langle{E_{\alpha}}\rangle}$) and initial 
energy distribution ($\Psi_{\alpha}(E)$).\\
In our analysis, neutrinosphere radius is taken as 10 km whereas other inputs
like initial flux and energy distribution depend on the choice of initial 
neutrino spectrum model. We analyze the evolution for several neutrino
spectrum model. 

\subsection{Models of Initial Neutrino Spectrum}

In a core-collapse SN, the gravitational binding energy (about a few times  
$10^{53}$ erg) is converted to neutrinos and antineutrinos with energies
of the 
order of 10 MeV and gets emitted in the subsequent $\sim 10$ sec. Initially a 
neutronization burst comes out consisting of pure $\nu_{e}$s but with only
a very small fraction of the total energy 
and after that the thermal neutrinos  
and antineutrinos of all three flavors are emitted. For the thermal neutrinos 
the initial energy distribution is expected to be Fermi-Dirac (FD), but the  
results of several simulations 
\cite{Totani:1997vj, Thompson:2003TBP, Keil:2002in} 
found that the distribution must be close to 
pinched thermal spectra \cite{Keil:2002in} i.e. with a deficit on the high 
energy side compared to FD.
Fermi-Dirac(FD) distribution in energy implies
\be
\Psi^{FD}_{\alpha}(E) \propto
\frac{\beta_{\alpha}~~(\beta_{\alpha} E)^{2}}{e^{\beta_{\alpha} E} + 1}~,
\ee
and for a choice of average energies of different flavors
\be
\langle E_{\nu_{e}} \rangle = 10~{\rm MeV}, 
~~\langle E_{\bar\nu_{e}} \rangle = 15~{\rm MeV},
~~\langle E_{\nu_{x}} \rangle = \langle E_{\bar\nu_{x}} 
\rangle = 24~{\rm MeV}, 
\ee
the inverse temperature parameters are \cite{Fogli:2007bk}
\be
\beta_{\nu_{e}} = 0.315~ {\rm MeV}^{-1},~~\beta_{\bar\nu_{e}} 
= 0.210~ {\rm MeV}^{-1},~~\beta_{\nu_{x}} 
= \beta_{\bar\nu_{x}} = 0.131 {\rm MeV}^{-1}.
\ee
Whereas the pinched spectra for different simulations are parameterized as 
\cite{Keil:2002in}
\be
\Psi_{\alpha}(E) =\frac{{\left(1+\zeta_{\alpha}\right)}^{1+\zeta_{\alpha}}}
{\Gamma(1+\zeta_{\alpha})}\left({\frac{E_{\alpha}}{\langle E_{\alpha}
      \rangle}}\right)^{\zeta_{\alpha}}\frac{\exp\left({-(1+\zeta_{\alpha})\frac{E_{\alpha}}
{\langle E_{\alpha} \rangle}}\right)}{\langle E_{\alpha}
      \rangle}~,
\label{eq:pinched}
\ee
$\langle E_{\alpha} \rangle$ is the average
energy of $\nu_{\alpha}$, and $\zeta_{\alpha}$ is the pinching
parameter. 
\\

\noindent
The effective number density for the $\alpha$th species per 
unit energy is given by
\be
n_{\alpha}(r,E) = \frac{D(r)}{2\pi R_{\alpha}^{2}}
\frac{L_{\alpha}}{\langle{E_{\alpha}}\rangle}\Psi_{\alpha}(E)~.
\ee
For a specific choice of $\Psi_{\alpha}(E)$ the 
initial flux ($\phi_{\alpha}$ = 
$\frac{L_{\alpha}}{\langle{E_{\alpha}}\rangle}$) for
the $\alpha$th flavor need to be specified and are 
very crucial input parameters
in our study.
Supernova models tell us that almost all the gravitational
energy released in core collapse supernovae comes 
out as $\nu \bar\nu$s of all
flavors. Only one or two percent of it goes into the explosion and the 
electromagnetic radiation emitted in all wavelengths.  
The total luminosity scales as $L(t)=L_{0} (e^{-t/\tau}/\tau)$ but for a
first study we take a time-averaged value for it as done in \cite{Fogli:2007bk} and \cite{Fogli:2009rd}.
One can of course look at the problem for specific instants of time by
changing the total luminosity, early times having larger values.

The total SN binding 
energy released ($ E_{B} $ = $3\times 10^{53}$ erg) is related to the 
individual flavor luminosities by
\be
L_{\nu_{e}} + L_{\bar\nu_{e}} + 4L_{\nu_{x}} = \frac{E_{B}}{\tau}~,
\ee 
assuming no distinction between $\nu_{x}$ and $\bar\nu_{x}$.
We also assume a time-independent constant luminosity over 
the time $\tau$. We take $\tau$ = 10
seconds. Thus the initial fluxes of different flavors get constrained by
\be
\phi^{0}_{\nu_{e}}\langle E_{\nu_{e}} \rangle + 
\phi^{0}_{\bar\nu_{e}}\langle E_{\bar\nu_{e}} \rangle 
+ 4\phi^{0}_{\nu_{x}}\langle E_{\nu_{x}} \rangle = 3\times 10^{52}~.
\ee
If we denote the ratio between the initial fluxes of different flavors by
\be
\phi^{0}_{\nu_{e}} : \phi^{0}_{\bar\nu_{e}} : \phi^{0}_{\nu_{x}} 
= \pre  : \prae  : 1~, 
\ee
where $\pre $, $\prae $ are positive numbers, then Eq. (20) can be written as
\be
\phi^{0}_{\nu_{x}}(\pre \langle E_{\nu_{e}} \rangle + \prae 
\langle E_{\bar\nu_{e}} \rangle + 4\langle E_{\nu_{x}} \rangle) 
= 3\times 10^{52}~.
\ee
Note that $\pre $ = {\large$\frac{\phi^0_{\nue}}{\phi^0_{\nu_x}}$ }, 
$\prae $ = {\large $\frac{\phi^0_{\anue}}{\phi^0_{{\nu}_x}}$} 
are basically initial relative fluxes.
Thus different choices of $\pre$ and $\prae$ imply
different relative luminosities or relative fluxes.

Four representative sets for the energy spectra (in terms of 
$\langle E_{\alpha} \rangle$ and the pinching factor $\zeta_{\nu}$)  and flux 
ratios usually discussed in literature, are given in Table 1.
One simulation by the Lawrence Livermore group (LL) \cite{Totani:1997vj} 
and two different 
simulations by the Garching group (G1, G2) \cite{Keil:2002in} are presented. 
Recently \cite{Dasgupta:2009mg} used another 
set of ``plausible'' flux parameters giving rise to 
multiple splits in the neutrino spectra is also given. 
We call this `G3'. For the LL spectra we use the FD distribution 
for $\Psi$ given in Eq. (14). The $\beta_\alpha$ for LL are 
given in Eq. (16). For G1, G2 and G3 spectra we use the pinched 
spectrum defined in Eq. (17). We assume 
$\zeta_{{\nu}_x}$=$\zeta_{\bar{\nu}_x}=4$ and $\zeta_\nue$=$\zeta_\anue=3$ for
G1 and G2. For G3 all $\zeta_{\alpha}= 3$.

\begin{table}[ht]
\begin{center}
\begin{tabular}{llllll}
\hline
Model & $\langle E_{\nue} \rangle$ & $\langle E_{\anue} \rangle$&
$\langle E_{\nu_x, {\bar{\nu}_x}} \rangle$ &$\pre$={\large 
$\frac{\phi^0_{\nue}}{\phi^0_{\nu_x}}$ } &$\prae$={\large 
$\frac{\phi^0_{\anue}}{\phi^0_{{\nu}_x}}$}\\
\hline
LL & 12 & 15 & 24 & 2.00 & 1.60 \\
G1 & 12 & 15 & 18 & 0.80 & 0.80 \\
G2 & 12 & 15 & 15 & 0.50 & 0.50 \\
G3 & 12 & 15 & 18 & 0.85 & 0.75 \\
\hline
\end{tabular}
\caption{The parameters of the used primary neutrino spectra models
  motivated from SN simulations of the Garching (G1, G2) and the
  Lawrence Livermore (LL) group. We assume 
$\zeta_{{\nu}_x}$=$\zeta_{\bar{\nu}_x}=4$ and $\zeta_\nue$=$\zeta_\anue=3$ for
G1 and G2. For G3 all $\zeta_{\alpha}= 3$. For LL we use a pure FD 
spectrum.}
\label{SNmodeltable1}
\end{center}
\end{table}
Note that the LL simulation obtained a large hierarchy $\langle
E_{\nue}\rangle<\langle {E}_{\anue}\rangle<\langle E_{\nu_x}\rangle\approx
\langle {E}_{\bar{\nu}_x}\rangle$, and an almost complete equipartition of
energy among the flavors. The Garching simulations predict a smaller hierarchy
between the average energies, incomplete equipartition, and increased spectral
pinching. The differences in the values of these parameters arise from the
different physics inputs.

\noindent
The equipartition of energy implies
\be
L_{\nu_{e}} = L_{\bar\nu_{e}} = L_{\nu_{x}}~. 
\ee 
In terms of our notation it means 
\be
\pre = \frac{\langle E_{\nu_{x}} \rangle}{\langle E_{\nu_{e}} 
\rangle} ;~~\prae=\frac{\langle E_{\nu_{x}} \rangle}
{\langle E_{\bar\nu_{e}} \rangle}~.
\ee
So complete equipartition for the Garching simulations would imply flux 
ratios (Table 2) different from the values in Table 1. Recent analyses 
\cite{Fogli:2009rd} have shown that the multiple 
split cases have origin in the departure from energy equipartition.\\
\begin{table}[ht]
\begin{center}
\begin{tabular}{llllll}
\hline
Model &$\pre$={\large 
$\frac{\phi^0_{\nue}}{\phi^0_{\nu_x}}$} & $\prae$={\large $\frac{\phi^0_{\anue}}{\phi^0_{{\nu}_x}}$}\\
\hline
G1 & 1.50 & 1.20 \\
G2 & 1.25 & 1.00 \\
G3 & 1.50 & 1.20 \\
\hline
\end{tabular}
\caption{The flux ratios for the Garching models with equipartition of energy}
\label{SNmodeltable2}
\end{center}
\end{table}
Actually there is no reason that equipartition should be 
strictly followed for the energy 
released from a real supernova.
In the next subsection we make extensive analysis of this multiple split 
phenomena with varying initial fluxes, which is equivalent to varying
$\pre$ and $\prae$.

\subsection{Survival Probability and Flux}

As stated above in this subsection we discuss the impact due to the variation 
of initial relative fluxes ($\pre$ and $\prae$) on the final spectrum. The
final spectrum is calculated at 400 km as collective effect is expected to
vanish at around 400 km. We also analyze this effect for different models of
initial neutrino spectrum 
spectrum LL, G1 and G3.

In principle the values of $\pre$ and $\prae$ can lie in a large range. 
Thus analysing this variation would require study in a wide range of 
the  $\pre$-$\prae$ parameter space.
Instead we consider the suggestion \cite{Lunardini-Smirnov-0302033}
that the uncertainty in the relative luminosities of different 
flavors must be 
in the range
\be
\frac{1}{2} \leq \frac{L_{\nu_{e}}}{L_{\nu_{x}}} \leq 
2~~~~;~~~~\frac{1}{2} \leq \frac{L_{\bar\nu_{e}}}{L_{\nu_{x}}} \leq 2~.
\ee
These limits in turn will put a constraint on the parameters $\pre$ and $\prae$
\be
\frac{1}{2} \frac{\langle E_{\nu_{x}} \rangle}{\langle E_{\nu_{e}} \rangle} \leq \pre \leq 2\frac{\langle E_{\nu_{x}} \rangle}{\langle E_{\nu_{e}} \rangle} ~~;~~\frac{1}{2} \frac{\langle E_{\nu_{x}} \rangle}{\langle E_{\bar\nu_{e}} \rangle} \leq \prae \leq 2\frac{\langle E_{\nu_{x}} \rangle}{\langle E_{\bar\nu_{e}} \rangle}~.
\ee
In Table 3 we present the lower limits (ll) and upper limits (ul) of the initial relative fluxes for different spectrum models LL, G1 and G3. 

\begin{table}[ht]
\begin{center}
\begin{tabular}{llllllll}
\hline
Model & $\langle E_{\nue} \rangle$ & $\langle E_{\anue} \rangle$&
$\langle E_{\nu_x, {\bar{\nu}_x}} \rangle$ &
{\large $ \phi^{r}_{\nu_{e; ll}}$ } &
{\large $ \phi^{r}_{\nu_{e; ul}}$ } &
{\large $ \phi^{r}_{\bar\nu_{e; ll}}$ } &
{\large $ \phi^{r}_{\bar\nu_{e; ul}}$ } \\
\hline
LL & 10 & 15 & 24 & 1.20 & 4.80 & 0.80 & 3.2 \\
G1 & 12 & 15 & 18 & 0.75 & 3.00 & 0.60 & 2.4 \\
G3 & 12 & 15 & 18 & 0.75 & 3.00 & 0.60 & 2.4 \\
\hline
\end{tabular}
\caption{The average energies, upper limits (ul) and lower limits (ll) of the
  initial relative flux for the models used.}
\label{SNmodeltable3}
\end{center}
\end{table}
To compare different flux models and study more of the parameter space we vary
$\pre$ and $\prae$ in the range [0.5,5.0] and [0.5,3.5] respectively,  
for all the models.
We find that varying $\pre$ and $\prae$ give rise to different possibilities of 
final spectra as discussed in \cite{Fogli:2009rd}. 
In addition to that we check it for different initial spectrum models. Here it
is notable that usually the initial spectrum models come with a fixed value of
$\pre$ and $\prae$ (see Table 1) but the main idea in this analysis is about
varying $\pre$ and $\prae$. So here by initial spectrum models (like LL,
G1, G3) we mean the energy dependence ($\zeta_{{\nu}_\alpha}$) and neutrino
average energies ($\langle E_{\nu_{\alpha}} \rangle$) of the models.
In what follows, we will see that for the 
Inverted Hierarchy (IH) the final spectrum is 
very sensitive to the values of $\pre$, $\prae$ and the model of initial
spectrum. Whereas for Normal Hierarchy (NH), the results are less dependent
on these quantities. We will discuss the reasons for this behavior.

\subsubsection{Probability and Flux: NH}
 
As already discussed in \cite{Dasgupta:2009mg, Fogli:2009rd}, large 
flux (luminosity) of $\nu_{x}$ can induce simultaneous swap in both 
neutrino and antineutrino sector for NH. 
In these cases initially the system is in an unstable equilibrium. As it  
evolves, it partially swaps the flavor in both neutrino as well 
as antineutrinos, to end up in a stable state. 
We further study this over different spectrum models and initial relative 
fluxes. 
For each of the different models we vary $\pre$ and $\prae$ in the range 
[0.5,5.0] and [0.5,3.5] respectively. We find 
that for several choices of ($\pre$,$\prae$) there is 
simultaneous swap in both neutrino and antineutrino spectrum 
and this swap may generate prominent split in the final 
spectrum. For a specific model these split energies 
($E_{c}$) may vary from low to
high energies, depending on the value of ($\pre$,$\prae$).

Independent of the choice of spectrum models, 
these split features are seen for 
low values of ($\pre$,$\prae$), which implies large flux of $\nu_{x}$
compared to other flavors \cite{Fogli:2009rd}. 
As the values of ($\pre$,$\prae$) increase, the split energy 
($E_{c}$) also increases, and close to the equipartition point the split
energy tends to infinity.

\begin{figure}
\vskip -1.5 cm
\includegraphics[width=6.0cm,height=8.5cm,angle=270]{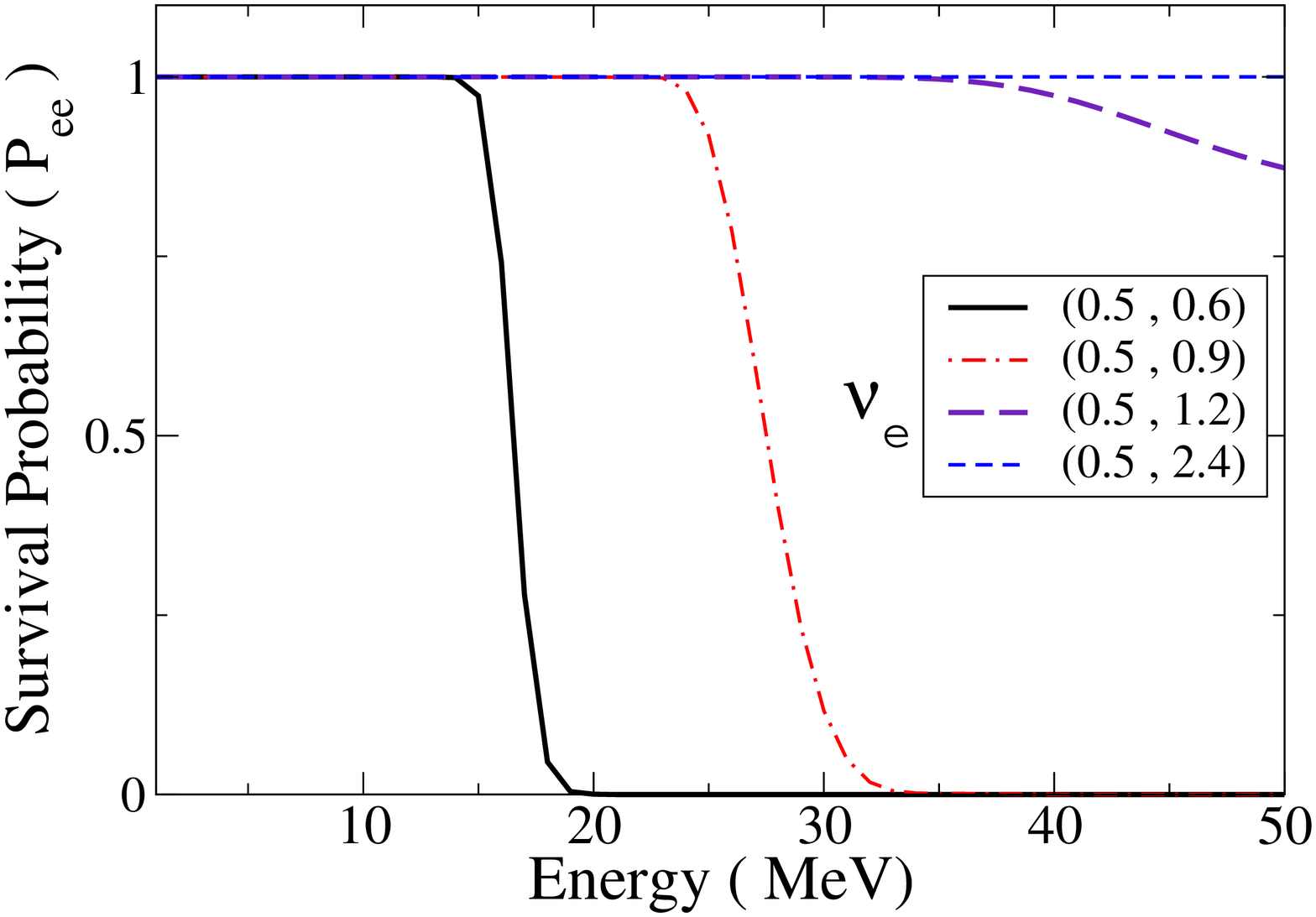}
\vglue -6.0cm \hglue 8.5cm
\includegraphics[width=6.0cm,height=8.5cm,angle=270]{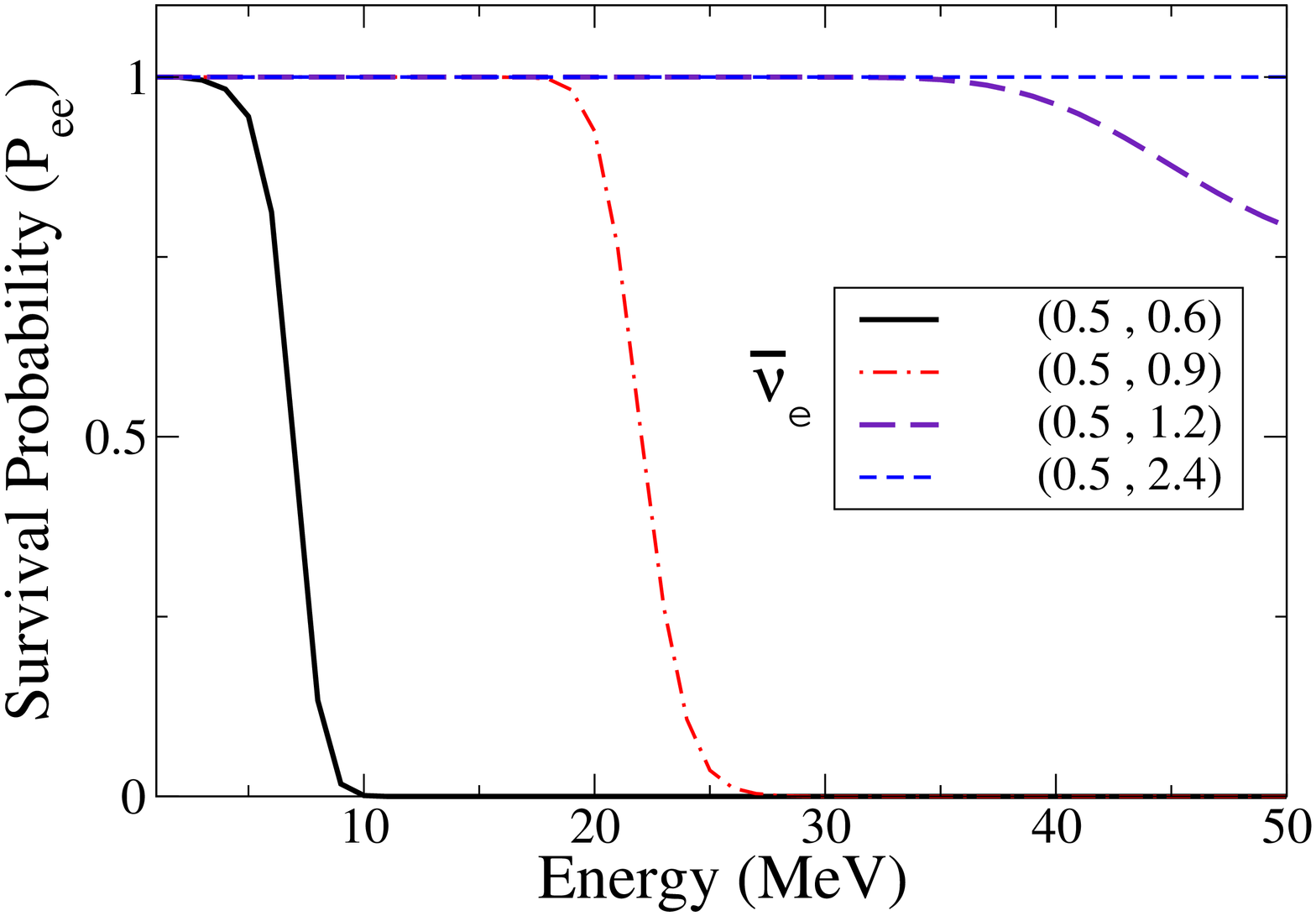}
\caption{\label{fig:flux}
\footnotesize{
Survival probability for G3 spectrum in 
NH with different ($\pre$,$\prae$).
}}
\end{figure}
\begin{figure}
\vskip -0.10 cm
\includegraphics[width=9.0cm,height=17.0cm,angle=270]{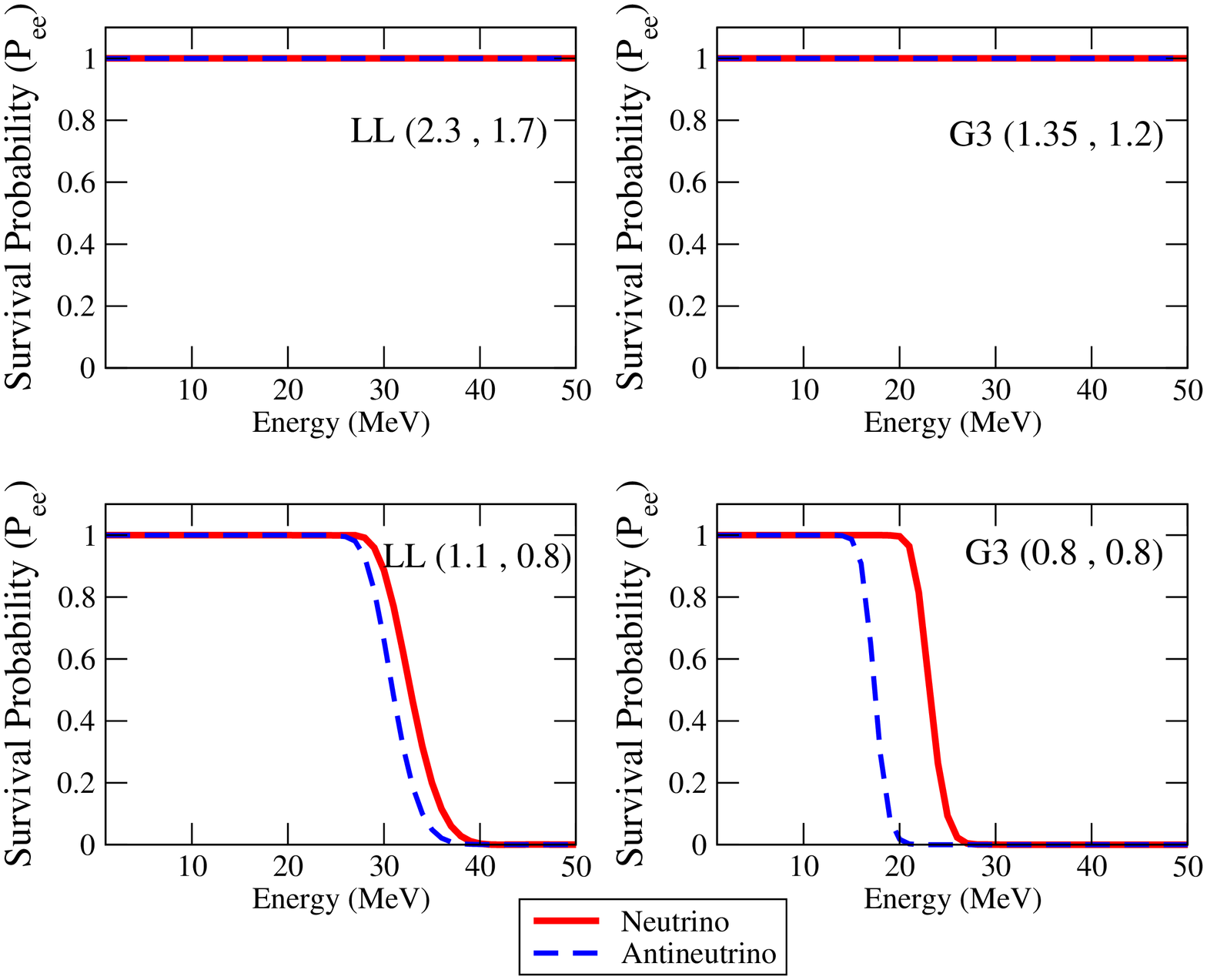}
\caption{\label{fig:probFDG3}
\footnotesize{
Survival probability for LL and G3 spectrum in NH with different ($\pre$,$\prae$).
}}
\end{figure}

As an example see Figure 1 where survival probabilities are plotted for the 
spectrum G3. The left panel is for neutrino and right one for antineutrino. 
For a low value of $\pre$ (0.5), $\prae$ is increased from 0.6 to 2.4 
for both neutrino and antineutrino. 
>From the figures it is evident that with this increment, split energies
($E_{c}$) also increase. 
For the same combination of $\pre$ and $\prae$ the split energy is higher for 
neutrinos than for antineutrinos. 
We find these features are same for other spectrum
models too. 

\begin{figure}
\vskip -0.3 cm
\hskip 1.0 cm
\includegraphics[width=4.5cm,height=8cm,angle=270]{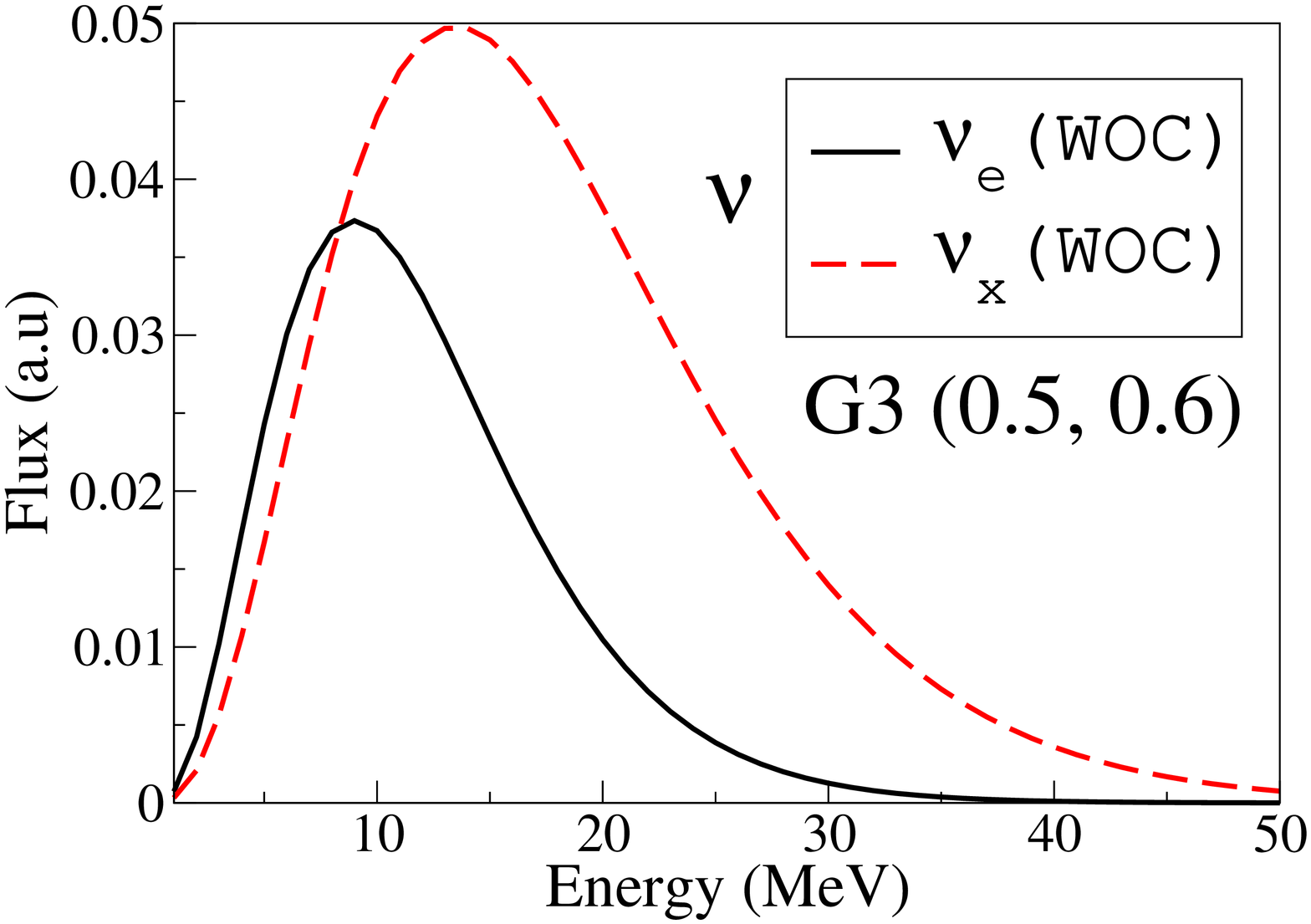}
\vglue -4.5cm 
\hglue 8.0cm
\hskip 0.5 cm
\includegraphics[width=4.5cm,height=8cm,angle=270]{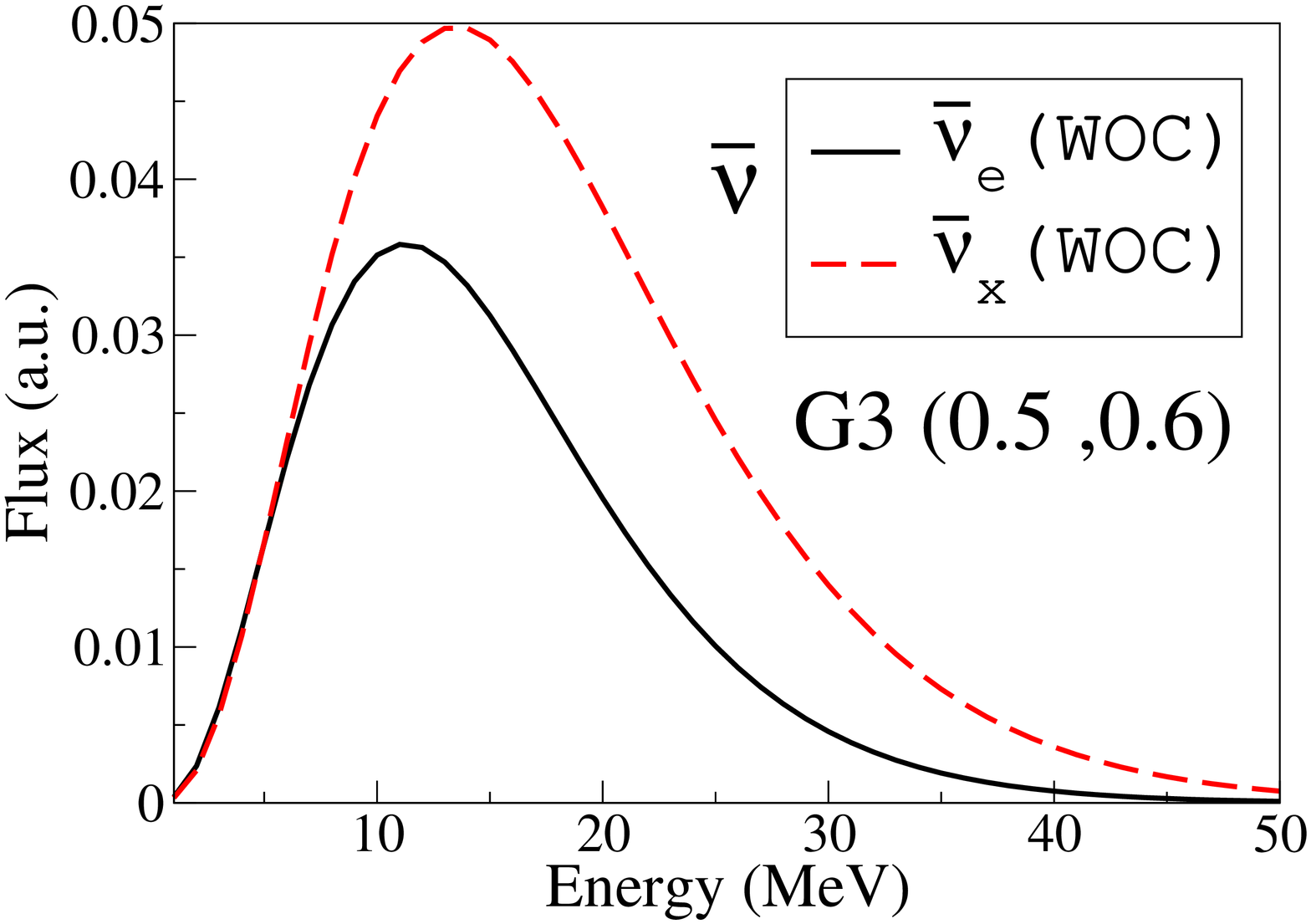}
\vskip -0.35 cm
\hglue 1.0cm
\includegraphics[width=4.5cm,height=8cm,angle=270]{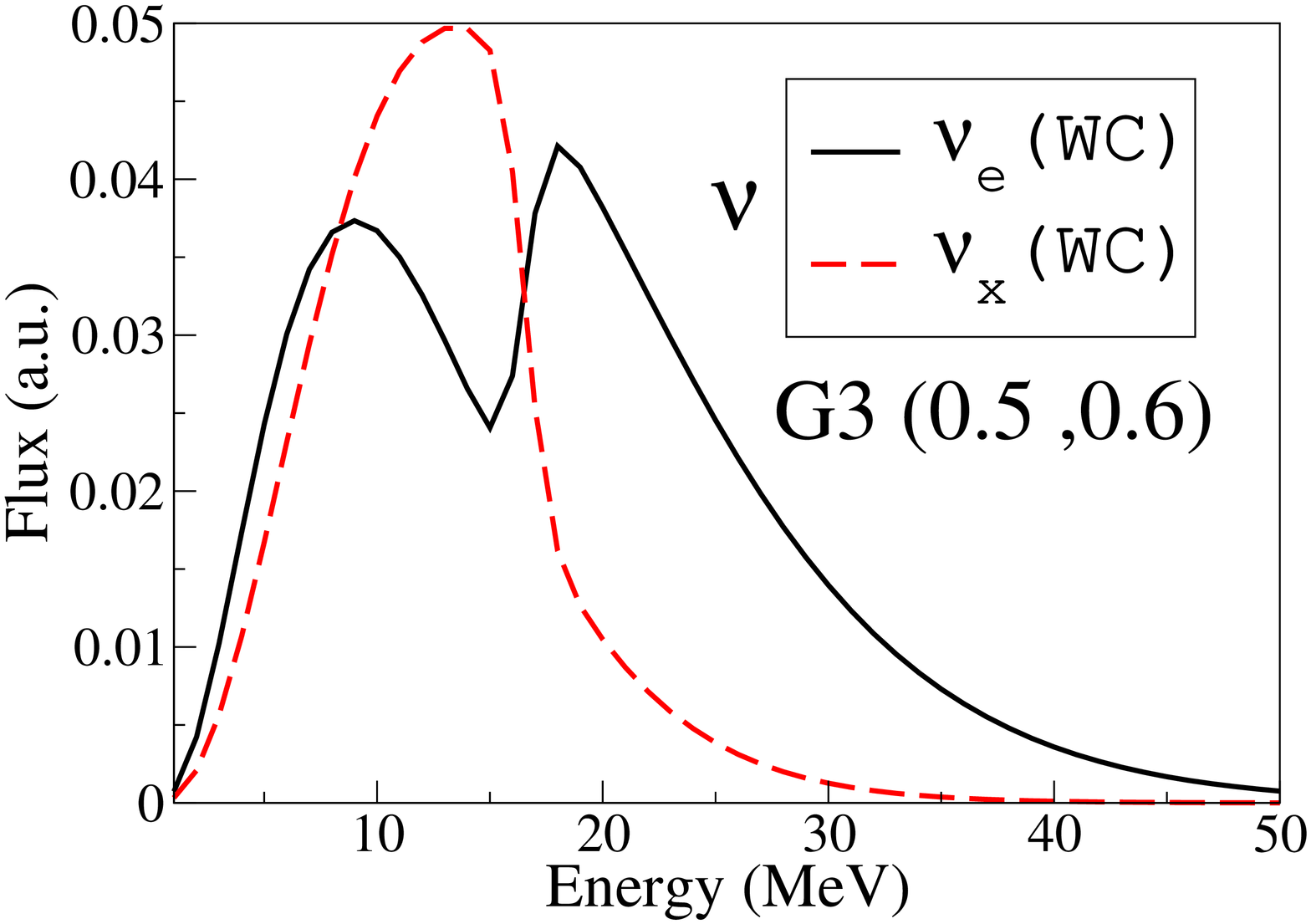}
\hskip -0.65 cm
\includegraphics[width=4.5cm,height=8cm,angle=270]{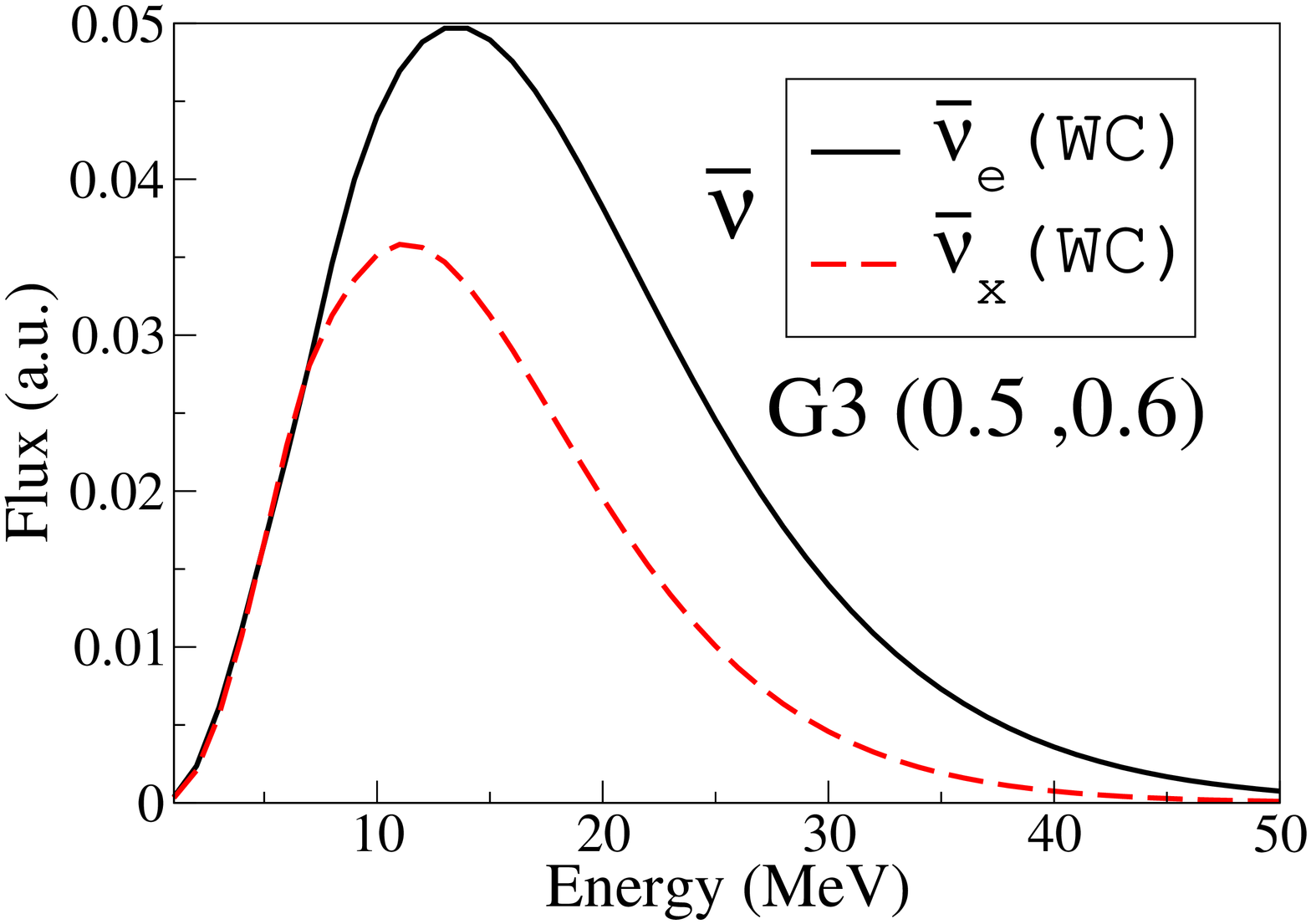}
\caption{\label{fig:flux2}
\footnotesize{
Flux for the different neutrino species for G3 with NH in arbitrary units (a.u.); 
WOC stands for ``Without 
Collective'' effects and WC for ``With Collective'' effects.}}
\end{figure}
In Figure 2  we have shown the above mentioned features 
for LL and G3. The left panels are for LL and the right ones are for G3. 
The red straight lines are for neutrino and the blue dashed lines are for 
antineutrino. In the top panels the initial relative spectrum ($\pre$,$\prae$)
are chosen to be close to the equipartition point for both the models and for 
these values there is no split whereas for the lower panels ($\pre$,$\prae$) 
are smaller and these panels show split for both neutrino and antineutrino
sectors.

The flux corresponding to G3 and (0.5,0.6) are plotted in Figure 3. 
Left and right panels in this figure 
are respectively for neutrinos and antineutrinos.  
The upper panels are without collective effects (initial flux) and the 
lower ones 
are with collective effect (flux beyond collective region). The black lines in
the panels are for electron type whereas the red dashed lines are for
x-type. Clearly the lower panels show swap in both neutrino and antineutrino
sector. The swap in both sectors are partial, that is, a part of the spectra
below the ``split energy'' remains same. The antineutrino split feature is not
clearly visible since $E_c$ for them is low and the $\anue$ and $\bar\nu_x$ 
fluxes are very close to each other at these energies. The
probability plots for (0.5,0.6) in Figure 1 also show the low split energy for
antineutrino. Note that the swap for neutrino spectra happens at a higher
energy compared to the antineutrino 
spectra, this feature is also consistent with the probability plots in
Figure 1.

\subsubsection{Probability and Flux: IH} 

Probability and flux in the IH is much more complex and interesting than NH.
Here also we vary the initial relative flux for different spectrum models and 
find wide variation of the final spectrum depending on the choice of 
($\pre$,$\prae$). 
These variations in spectrum with initial relative
flux ($\pre$,$\prae$) have been attributed to meeting the instability
condition of the initial system and adiabaticity violation
\cite{Dasgupta:2009mg} as well as to change of global initial condition with
luminosity variation
and minimization of potential energy \cite{Fogli:2009rd}. We find that these
changes in final spectrum are similar for different choices of spectrum
models.

As discussed in \cite{Fogli:2009rd}, the different spectral features arise from
the initial conditions, which may or may not lead the system to  
swap to minimize ``potential energy''. We also find that in some cases 
the multiple swaps actually do take place but the 
swaps are so close that they can not 
be resolved numerically
\cite{Dasgupta:2009mg} and thus appear as if the swap or split features are 
absent.
\begin{figure}
\vskip -0.5 cm
\includegraphics[width=10.0cm,height=19.0cm,angle=270]{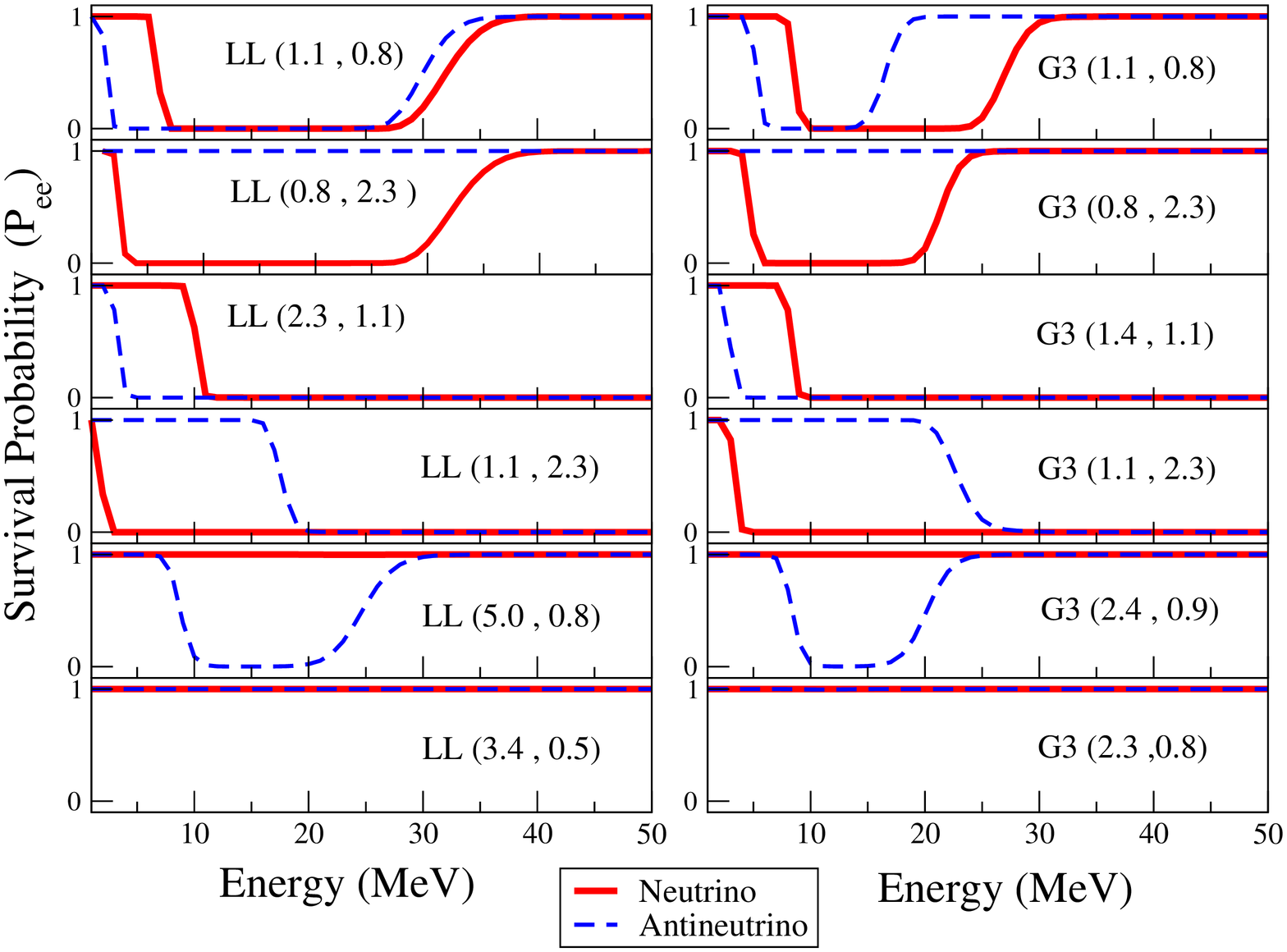}

\caption{\label{fig:probFDG31}
\footnotesize{
Survival Probability for LL and G3 spectrum in IH with different ($\pre$, $\prae$).
}}
\end{figure}
We find five spectral split patterns
as mentioned in \cite{Fogli:2009rd}. 
These five patterns are found for all three models of initial energy 
spectra LL, G1, G3.  
These  are displayed (for LL and G3) in successive panels from top to bottom 
in Figure \ref{fig:probFDG31}.
\begin{enumerate}
\item
Dual split in both neutrino and antineutrino flux (II,II). 
\item
Dual split in neutrino but no split in antineutrino flux  (II,0).
\item
One split in both neutrino and antineutrino flux with the split energy of the  
neutrino higher (H) than that of antineutrino (L) split energy (I,I)(H,L). 
\item
One split in both neutrino and antineutrino flux with 
the split energy of the neutrino lower (L) than  that of antineutrino (H) split energy (I,I)(L,H).
\item
No split in neutrino but dual split in antineutrino flux (0,II). 
\end{enumerate}

Apart from these five patterns we find a sixth possible pattern in
which neither neutrino nor antineutrino show any swap in the spectrum. 
We call this (0,0). For this pattern (0,0) the effect of neutrino-neutrino
interaction on both neutrino and anti neutrino flux is undetectable. 

The physical reasoning behind the patterns in the top five panels are well 
explained \cite{Fogli:2009rd} from the idea of
potential energy minimization. Our analysis shows that in some sense all the
different spectrum models are in the same footing as all of them give rise to 
similar split patterns with the change of initial relative flux or relative 
luminosity.
As explained in \cite{Dasgupta:2009mg} the basic feature is that there are 
multiple swaps or splits in both the neutrino and antineutrino sector but the swaps may 
disappear depending on the adiabaticity violation or it may be numerically 
unresolvable.split patterns
Consider the new pattern, 
described in the lowest panel of Figure 4, where it seems that 
there is no swap 
in both neutrino and antineutrino sectors. 
\begin{figure}
\includegraphics[width=6.0cm,height=19.0cm,angle=270]{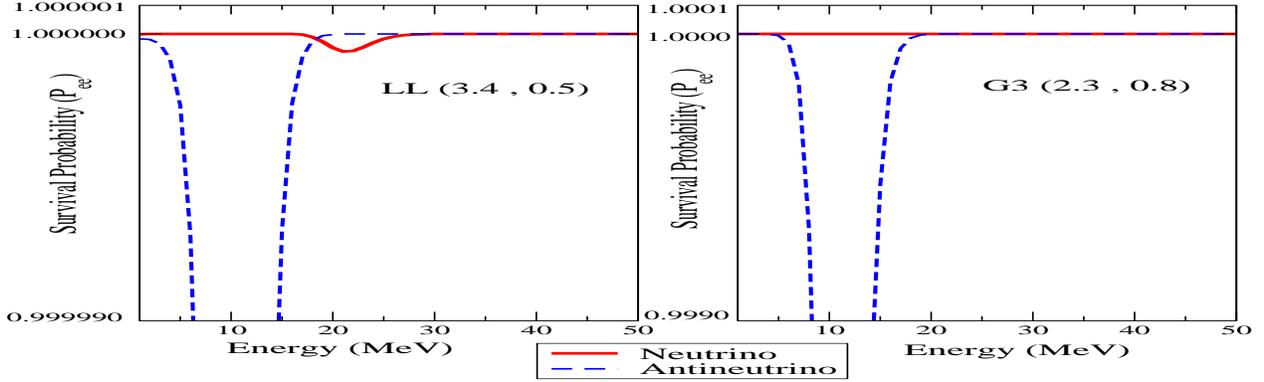}

\caption{\label{fig:probLLG3}
\footnotesize{
 Survival Probability for LL (3.4, 0.5) and G3 (2.3, 0.8) spectrum in IH .
}}
\end{figure}
When we study these cases carefully (Figure 5, left panel LL (3.4,0.5) and right panel G3 (2.3,0.8)) we find that they also show changes in survival probability similar to the other patterns. But the swaps here are incomplete and numerically undetectable. While in Figure 4 the change in probability for this case is 
visually unresolvable for all practical purposes, 
in Figure 5 it is visible, as 
we have increased the resolution.

For the fluxes we just give one example of the case (II,II) in Figure 6. 
Here we plotted the G3 neutrino spectrum in the left panel and the G3
antineutrino spectrum in right one. In both 
panels, the solid sky blue lines are for electron type
without collective effect (WOC) and the solid 
red lines are for $\nu_x$ without 
collective effect (WOC). For the spectrum with collective effects (WC) dashed
black lines are for electron type whereas dot-dashed blue lines are for 
$\nu_x$. Here the spectrum model 
used is G3 and the initial relative 
fluxes  are (1.1,0.8). We can see prominent 
dual split pattern in this flux figure as expected from the upper 
right panel of Figure 4.
\begin{figure}
\includegraphics[width=7cm,height=8.9cm,angle=270]{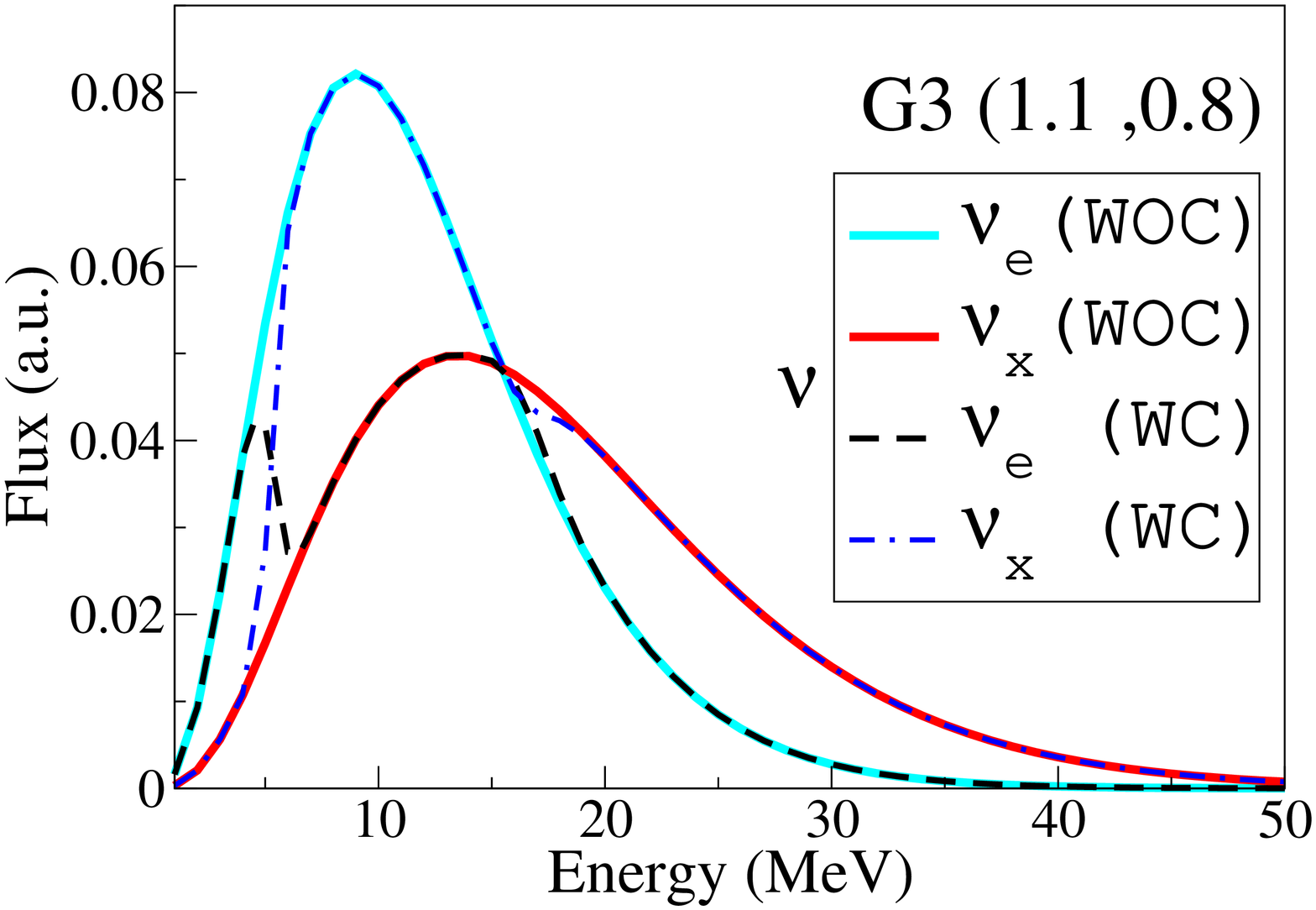}
\hglue -0.50cm
\includegraphics[width=7cm,height=8.9cm,angle=270]{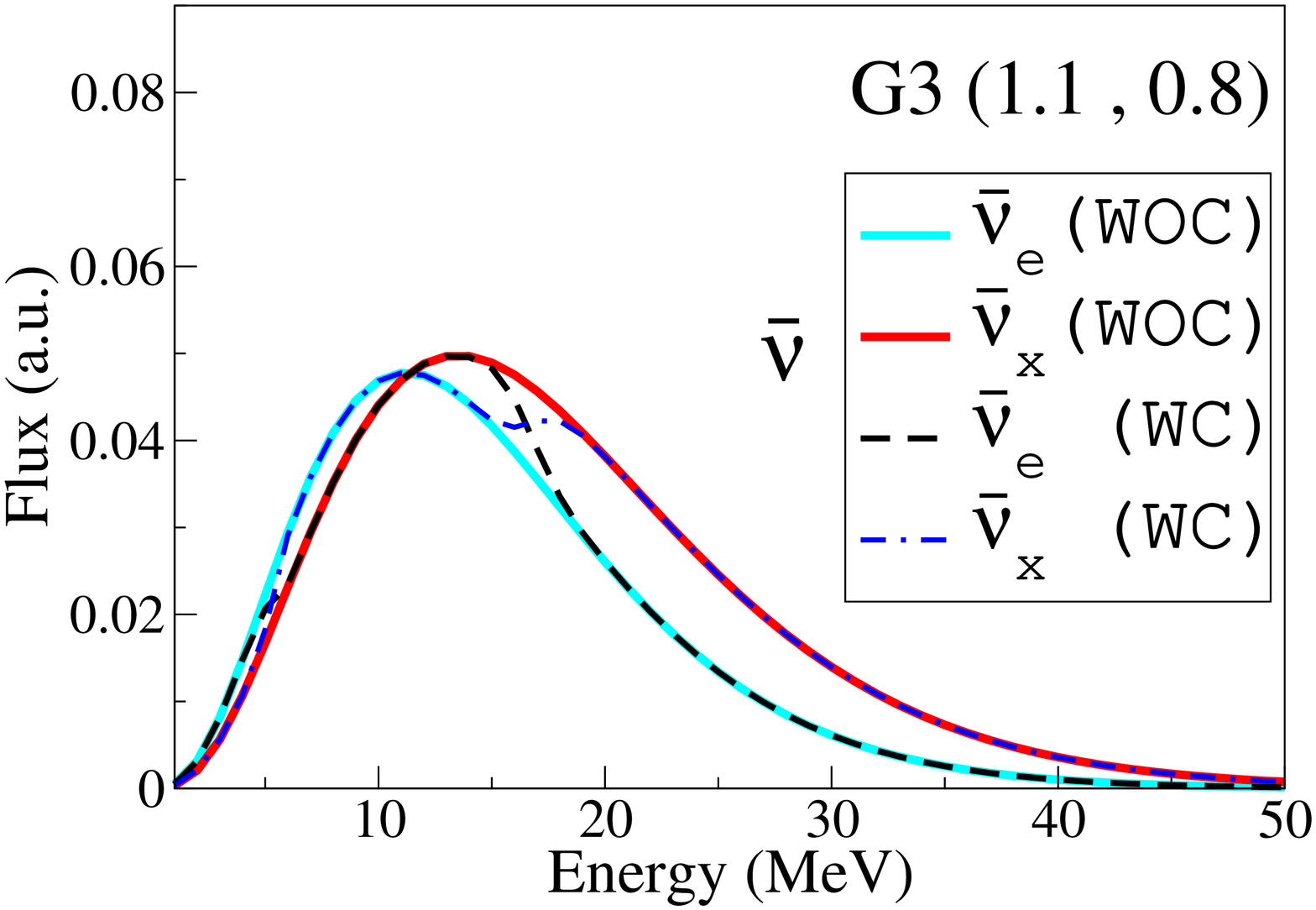}
\caption{\label{fig:flux0}
\footnotesize{
The neutrino and antineutrino fluxes in arbitrary units (a.u.) for the model G3 with the relative 
luminosities (1.1,0.8),  with and without collective effect for 
IH. }
}
\end{figure}

Thus, depending upon the choice of initial relative fluxes
($\pre$,$\prae$) the 
spectra can have different patterns, especially for IH.  
The possible values of ($\pre$,$\prae$) can be in a wide range. 
Even if one assumes a factor-of-two-uncertainty in the relative luminosity
\cite{Lunardini-Smirnov-0302033},
there can be considerable variations in the final flux characteristics.

\begin{figure}
\includegraphics[width=10cm,height=19cm,angle=270]{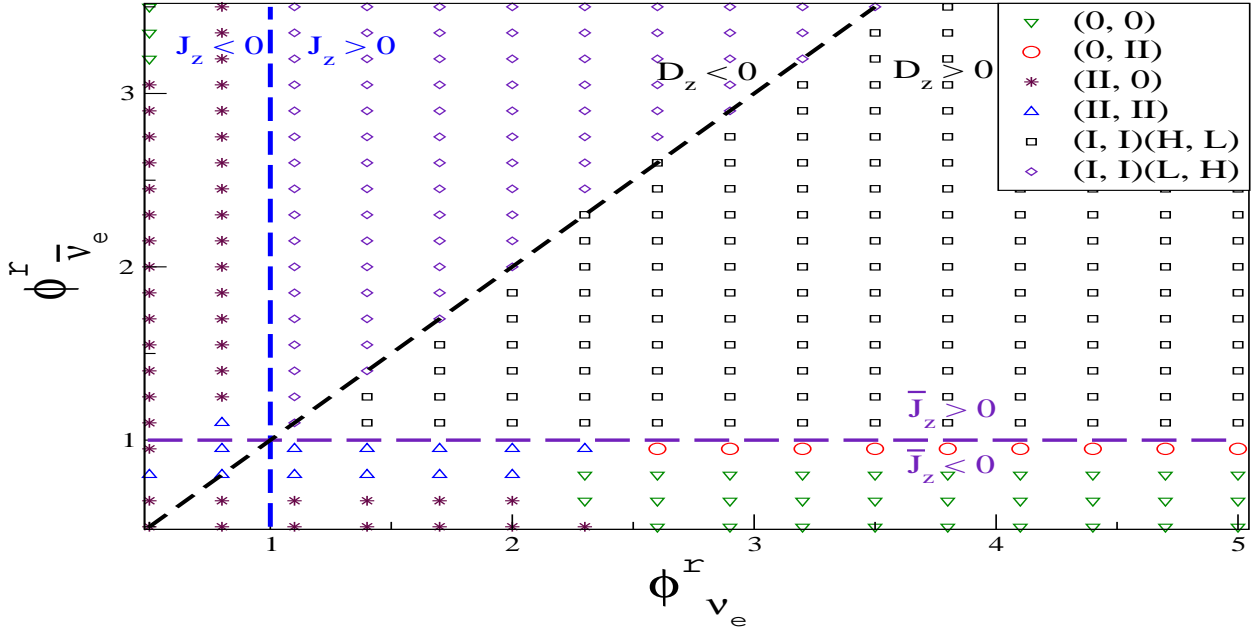}
\caption{\label{fig:Split-region}
\footnotesize{
The different split pattern regions in the $\pre$-$\prae$ plane for IH and G3. The black dashed, purple long dashed, blue thick dashed denotes  $D_z = 0$, $\bar J_z=0$ and $J_z=0$ lines respectively. The six zones in $\pre$-$\prae$ plane have different split patterns. The abbreviations used are explained in section 2.3.2. }
}
\end{figure}

We study the variation
in spectral split features over the $\pre$-$\prae$ plane for LL, G1, G3
and found a pattern showing different kind of spectral splits at different
$\pre$-$\prae$ region. 
In Figure \ref{fig:Split-region} we show this in the $\pre$-$\prae$ plane. 
The plane is divided into zones by the values of the 
global polarization vectors $J_z, \bar J_z$ and $D_z$. 
The black dashed line divides the plane into zones with 
$D_z>0$ and $D_z<0$. The purple long dashed corresponds to $\bar J_z=0$ and 
demarcates the area which has $\bar J_z$ positive and negative. 
The blue thick dashed  is for $J_z=0$. These lines therefore divide 
the $\pre$-$\prae$ plane into 6 zones. The split patterns observed in 
the different zones are shown on the plane.
The global polarization vectors $J_z, \bar J_z$ and $D_z$, define the ``phase transitions'' across different split patterns.
It should be noted that the global polarization vectors were initially in the
z direction hence the sign changes of their z components mark the stability of
the system and spectral splits \cite{Fogli:2009rd}. 
Figure  \ref{fig:Split-region} shows that 
\begin{enumerate}
\item
(I,I)(H,L) patterns are for $J_z > 0, \bar J_z > 0 $ and $D_z > 0 $,
\item
(I,I)(L,H) for $J_z > 0, \bar J_z > 0 $ and $D_z < 0 $,
\item
\begin{enumerate}
\item
(II,II) patterns are seen in the $J_z < 0, \bar J_z < 0 $ region, 
and also in the $J_z > 0, \bar J_z < 0 $ region,
\item
(0,0) appear mostly in $\bar J_z < 0 $ with a very few occurrence in  $J_z < 0$,
\item
(II,0) is the most dominant pattern in the $J_z < 0$ region, although it can appear in the $J_z > 0$, $\bar J_z < 0$ region,
\item
(0,II) pattern occurs only in $\bar J_z < 0 $.
\end{enumerate}
\end{enumerate}
Thus for the double split patterns described in point 3 above, the so called ``phase transition'' lines seems inconclusive. \cite{Fogli:2009rd} established that the $J_z < 0$ or $\bar J_z < 0 $ region (i.e, the rectangle covering the zone with $J_z <0$ alongwith the rectangle covering the zone with $\bar J_z < 0$ ) lead to (II,II) pattern when the adiabaticity is increased artificially. Hence due to the incomplete adiabaticity in the actual case some of the (II,II) patterns appear as (II,0), (0,II), (0,0) pattern. So with complete adiabaticity one will not see any of the (II,0), (0,II), (0,0) patterns in Figure \ref{fig:Split-region}. For example see the discussion regarding the (0,0) pattern in context of Figure \ref{fig:probFDG31}.

With so many possible    
patterns it will be really difficult to predict the initial relative neutrino fluxes, the energy distribution model and the extent of collective neutrino effect, even for a future galactic supernova event.In the next section we discuss to what extent one can constrain the luminosities by demanding a neutron rich condition by the end of the collective region, 
required for successful r-process. 

\section{Neutrino Fluxes and r-Process Nucleosynthesis}


\begin{figure}[ht]
\includegraphics[width=11cm,height=18cm,angle=270]{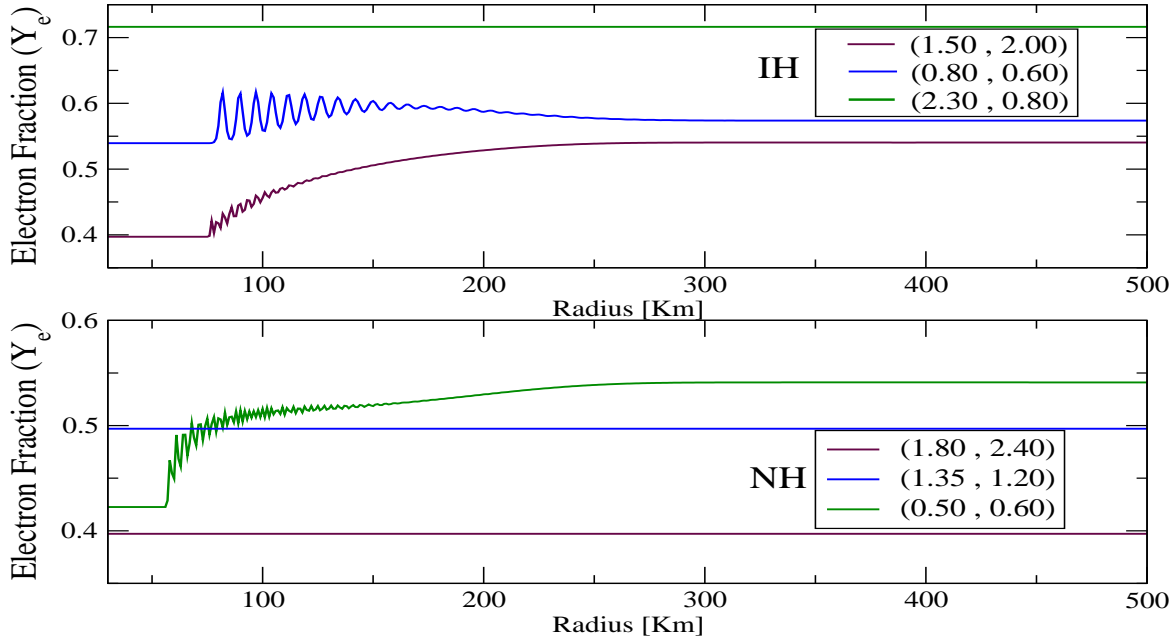}
\caption{\label{Y-evsr}
\footnotesize{
The electron fraction $Y_{e}$ for both hierarchies as a 
function of `r', the distance of the region 
from the center of the core.The spectrum model used is G3.}}
\end{figure}

In this section we discuss the effect of the flux of neutrinos radiated out
in core collapse supernovae on the electron fraction and discuss the 
possibility of getting allowed regions for r-process nucleosynthesis
and the resulting constraints on relative luminosities. 
As most simulations of core-collapse supernovae do not lead to explosions,
there are uncertainties in the understanding of the late stage of the SN shock 
propagation. But the generally accepted scenario supported by simulations
is that for  
core collapse supernovae starting with iron cores
the shock wave gets initially stalled due to
loss of energy through nuclear dissociation and then over timescale of a
second, gets revived by the energy deposited by neutrinos radiating out, the
so-called late-time neutrino heating mechanism 
leading
to the delayed core collapse supernova
\cite{Arnould}. 
This results in the
development of a low density ``hot bubble'' region just behind the SN shock.
Normally the hot bubble regions are
taken between the infalling neutron star radius and the forward
shock, that is, up to 30-40 km initially. 
The huge flux of neutrinos emitted from the proto-neutron star leads to
the ``neutrino-driven wind'' which remains active for about 10 seconds after
the core bounce. This creates neutron-rich regions of high entropy which
are conducive to the development of the r-process. 
Different 
delayed core-collapse SN calculations give rise to different values for the
entropy per baryon leading to conflicting conclusions about the r-process.
However the $\nu$-driven 
wind is still considered to be one of the most probable
sites for the r-process \cite{Arnould}. 
The $\nu$-driven wind models 
consider the r-process site to be at a few 
hundred kilometers (within 
1000 km) \cite{Arnould}. Since, as discussed before, 
this is also the region where 
collective oscillations are active, 
it is expected that r-process will get affected. 
In what follows, we study this effect 
in the $\nu$-driven wind region and numerically check if 
criteria of successful r-process can be used to constrain initial 
neutrino flux parameters.

The criteria for r-process on which we focus here is the 
the electron fraction $Y_e$, {\it i.e.}, 
the number of electrons (equal to the number of
protons, due to charge neutrality) per baryon. 
The $Y_e$ will 
depend on the relative 
strengths of the two reactions -- 
neutrino capture on neutrons and antineutrino
capture on protons.
Therefore, $Y_{e}$ can be expressed as
\cite{Qian:1993dg}
\be
Y_{e}=1/(1+\lambda_{\bar\nu_{e} p}/\lambda_{\nu_{e} n})~,
\label{eq:ye0}
\ee
where $\lambda_{\nue n}$ and $\lambda_{\anue p}$ are the reaction rates
for 
$\nu_e +n \rightarrow e^- + p $ and $\bar{\nu}_e + p \rightarrow e^+ + n$ 
respectively. 
Note that these reactions can in principle occur on both free and bound
nucleons. However for the purpose of this work, we will not consider the
reactions on heavy nuclei\footnote{For a detailed study of the effect of
nuclear compositions on $Y_e$ we 
refer to \cite{McLaughlin:1997qi}.
Our main conclusions come from impact of collective oscillations 
on r-process nucleosynthesis and 
are not expected 
to drastically change as a result of reactions on bound nucleons.}. 
Note also that 
in principle, the inverse reactions also happen inside 
the supernova and should be considered. However, 
we neglect the inverse reactions here since the matter temperature of the
region is small compared to the neutrino temperature as one goes away from
the neutrinosphere and has very small 
effect at radius of 30 km and beyond \cite{Qian:1993dg,Bethe:85}.
The reaction rates $\lambda_{\nu N}$ (where {$N = n$ or $p$}) are given 
as  
\be
\lambda_{\nu N} \approx \frac{L_\nu}
{4\pi r^2} \frac{\int_{0}^{\infty} 
\sigma_{\nu N}(E)f_\nu(E)dE}{\int _{0}^{\infty} E
f_{\nu}(E)dE}~,
\label{eq:lambda}
\ee
where $L_\nu=\phi_\nu^0\langle E_\nu\rangle$, and 
$f_\nu$ denotes the neutrino flux. 
The cross section used are
\be
\sigma_{\nu_{e} n}(E_{\nu_{e}}) \approx 9.6\times 10^{-44} \left(
  \frac{E_{\nu_{e}}+ \triangle_{np}}{MeV}\right)^{2} {\rm cm}^{2}~, 
\ee
\be
\sigma_{\bar\nu_{e} p}(E_{\bar\nu_{e}}) \approx 9.6\times 10^{-44} \left(
  \frac{E_{\bar\nu_{e}}- \triangle_{np}}{MeV}\right)^{2} {\rm cm}^{2}~,
\ee
where $\triangle_{np}=1.293$ MeV  
is the mass difference between neutron and proton. 

Note that the neutrino flux denoted as $f_\nu$ in Eq. (\ref{eq:lambda}) 
is the flux including collective flavor oscillations. 
The swap between the active neutrinos
due to collective effect can 
change $f_\nu$ and hence $Y_e$. To illustrate this better we show 
the ratio of the reaction rates for $\anue$ and $\nue$  
explicitly in terms of the collective oscillation probabilities 
\be
\frac{\lambda_{\bar\nu_{e} p}}{\lambda_{\nu_{e} n}}(r)=\frac{\int
  _{0}^{\infty} \sigma_{\bar\nu_{e}
    p}(E)P_{\bar\nu_{e}}^{c}(r,E)\prae\Psi_{\bar\nu_{e}}(E)dE+\int
  _{0}^{\infty} \sigma_{\bar\nu_{e}
    p}(E)(1-P_{\bar\nu_{e}}^{c}(r,E))\Psi_{\nu_{x}}(E)dE}{\int _{0}^{\infty}
  \sigma_{\nu_{e} n}(E)P_{\nu_{e}}^{c}(r,E)\pre\Psi_{\nu_{e}}(E)dE+\int
  _{0}^{\infty} \sigma_{\nu_{e}
    n}(E)(1-P_{\nu_{e}}^{c}(r,E))\Psi_{\nu_{x}}(E)dE}~,
\label{eq:ratio}
\ee
where $P_{\bar\nu_{e}}^{c}(r,E)$ and $P_{\nu_{e}}^{c}(r,E)$ are the anti-neutrino 
and neutrino survival probabilities with collective oscillations and are
calculated numerically as function of radius and energy. 
The minimal condition for the SN environment to become neutron reach is
$Y_e < 0.5$ which translate as the condition 
$\lambda_{\bar\nu_{e} p}/\lambda_{\nu_{e} n} > 1$. 

Let us begin by understanding the impact of collective flavor 
oscillations on r-process by discussing some limiting cases. 
n the no flavor oscillation limit, i.e. 
$P_{\bar\nu_{e}}^{c} = P_{\nu_{e}}^{c} = 1.0$  
for all energies, Eq. (\ref{eq:ratio}) reduces to 
\be
\bigg(\frac{\lambda_{\bar\nu_{e} p}}{\lambda_{\nu_{e} n}} \bigg)_{\rm no~osc}
\simeq \frac{\prae}{\pre} 
\frac{\int _{0}^{\infty} (E-\Delta_{np})^2 \Psi_{\bar\nu_{e}}(E)dE}
{\int _{0}^{\infty} (E+\Delta_{np})^2 \Psi_{\nu_{e}}(E)dE}
\simeq \frac{\prae}{\pre} 
\frac{\langle  (E-\Delta_{np})^2 \rangle _{\bar\nu_e}}
{\langle (E+\Delta_{np})^2 \rangle_{\nu_e}}
\label{eq:nooscratio}
\ee
Since the average energy of $\anue$ is greater 
than that of $\nue$ for all the three SN models that we have 
considered, it is expected that 
 $\langle  (E-\Delta_{np})^2 \rangle _{\bar\nu_e} >  
\langle (E+\Delta_{np})^2 \rangle_{\nu_e}$. 
Therefore under this approximation, 
for $\prae/\pre \geq 1$, $Y_e < 0.5$ always and r-process 
can proceed. 
The condition $Y_e\leq 0.5$ in fact gives $\prae/\pre \geq $   
0.62 for LL and $\geq $  0.88 for G1/G3 .
For all values of $\prae/\pre$ greater than this value, r-process 
can happen while for all values of $\prae/\pre$ below this, 
r-process is forbidden. 

Likewise one could consider the case where we have 
complete conversion of both neutrinos and antineutrinos where 
$P_{\bar\nu_{e}}^{c} = P_{\nu_{e}}^{c} = 0 $ for
 all energies.
One can easily show that for this case 
$\lambda_{\bar\nu_{e} p}/\lambda_{\nu_{e} n} = \langle (E-\triangle_{np})^2 \rangle_{\nu_x} / \langle (E+\triangle_{np})^2 \rangle_{\nu_x} $.
As a result here one always gets $Y_e > 0.5$ as $\triangle_{np}$ is positive. 
Note however that this case never happens 
in collective oscillations and is therefore not realistic. 

Next we consider effect of collective effects on r-process for the realistic 
case, where the flavor conversions are calculated numerically, 
as outlined in the previous section. 
In Figure \ref{Y-evsr} we show the electron fraction $Y_e$ as a 
function of the radius $ (r) $, for different combinations 
of $\pre$ and $\prae$. We have taken the G3 model for the 
$\nue$ and $\anue$ spectra. 
The upper panel is for IH while 
the lower one is for NH. 
In the upper panel, the green line is for 
($\pre$, $\prae$) of (1.5, 2.0), the blue line for 
(0.8, 0.6), while the maroon line is for (2.3, 0.8). 
We can note from the figure that for IH and (2.3, 0.8) case, there 
is no flavor conversion due to 
oscillations for the inverted hierarchy. 
The probability $P_{\bar\nu_{e}}^{c}$ and 
$P_{\nu_{e}}^{c}$ for this case was 
shown in the lowest right-hand panel of Figure \ref{fig:probFDG31}, 
where we can see that 
$P_{\bar\nu_{e}}^{c} = P_{\nu_{e}}^{c} = 1 $. Therefore  
as discussed above, the 
$\lambda_{\bar\nu_{e} p}/\lambda_{\nu_{e} n} = (0.8/2.3)*
\langle (E-\triangle_{np})^2 \rangle_{\bar\nu_e} / \langle (E+\triangle_{np})^2 \rangle_{\nu_e} = 0.396$, 
giving $Y_e\sim0.72$. For the two other cases considered with IH we have oscillations due to single and multiple splits. A scan of Figure \ref{fig:Split-region} reveals that we have 
double splits in both neutrino and antineutrino channels for the blue line with (0.80, 0.60) whereas for the maroon lines of (1.50, 2.00) we have single splits in both neutrino and antineutrino channels with the split energy of neutrino lower than that of antineutrino. For both these cases we can see very fast oscillations 
in $Y_e$ within the first 200 km, which can be attributed
to the bipolar collective oscillations.
Beyond 300 km the value of $Y_e$ approaches a fixed value 
as the neutrino density decreases very fast and the 
collective effects end. The reason that one 
gets higher values of $Y_e$ for
 both of them after the completion of collective effects 
compared to their values at 30 km can become clear from Eq. 
(\ref{eq:ratio}).
For double splits for fixed value of $\prae$/$\pre$ the contribution 
from the integrals in denominator in between the split 
energies is more than the corresponding contribution 
in the numerator making
the ratio lower, resulting in higher $Y_e$. On the other hand for the single split case of (1.50, 2.00) the low split energies in the denominator for neutrinos make the ratio of Eq. 
(\ref{eq:ratio}) lower.

In the lower panels we assume NH and show $Y_e$ for 
(0.5, 0.6) by the green  line, for (1.35, 1.2) by the 
blue line, and (1.8, 2.4) by the maroon line. 
For the (1.35, 1.2) case, we had noted before in Figure 
\ref{fig:probFDG3}, that $P_{\bar\nu_{e}}^{c} = P_{\nu_{e}}^{c} = 1$ 
over the entire energy range. For the case (1.8, 2.4) 
there is no conversion as well. 
Hence for these case there is no flavor conversion and $Y_e$ 
stays constant for all $r$, given solely in terms of the $\prae$/$\pre$ 
and
$ \langle (E-\triangle_{np})^2 \rangle_{\bar\nu_e} / 
\langle (E+\triangle_{np})^2 \rangle_{\nu_e}$ 
ratios. 
For the case (0.5, 0.6) (cf. Figure \ref{fig:flux}) we have 
single split in both the neutrino and antineutrino channels.  
For this case therefore we see a variation in $Y_e$ as a function 
of the radius. 
Since we have noted that for cases of ($\pre$, $\prae$) for 
which there is flavor conversion due to collective effects, 
$Y_e$ fluctuates non-trivially with the radius for $ r \ltap 400$ km,  
therefore in what follows, we will show all results for 
$r \gtap 400$ km. That implies that we consider only 
the neutrino-driven wind region henceforth. 

\begin{figure}[ht]
\vskip -1.5cm
\includegraphics[width=10.5cm,height=18cm,angle=270]{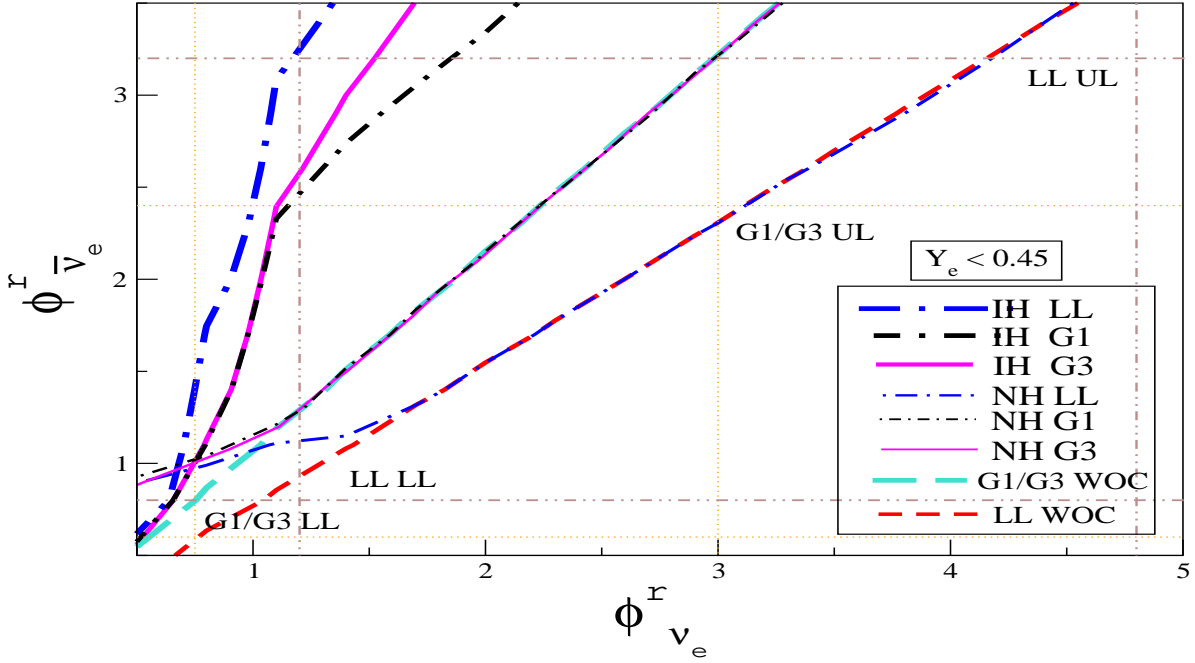}

\caption{
\label{Ye<0.45G13LLexcl}
\footnotesize{
 The exclusion plot consistent with $Y_{e}$ $<$ 0.45 for the spectrum G1, G3
 and LL for both NH and IH. The allowed area is to the left of the curves. The
 dotted orange lines denote the  Lower Limit (LL) and Upper
 Limit (UL) of $\pre$ and $\prae$ for  G1 and G3, arising from the
 two fold uncertainty defined in Eq. (25). Similarly double dotted dashed brown
 lines denote the  Lower Limit (LL) and Upper Limit (UL)
 for the Lawerence Livermore spectrum model (LL).}}
\end{figure}

Figure  \ref{Y-evsr} also shows the most significant aspect of 
flavor conversion due to collective effects. 
At $r=30$ km, one is in the synchronization phase and 
the collective bipolar oscillations are yet to set in. 
This is the limiting case of no conversion already discussed, 
while by $r=400$ km,  they are complete. 
The maroon line (1.5, 2.0) for IH and green 
line (0.5, 0.6) for NH show that collective effects can 
change the value of the electron fraction from $Y_e < 0.5$ 
(at $r=30$ km) 
to $Y_e > 0.5$ (at $r=400$ km). Hence, we can 
explicitly see that these combination of values of $\pre$ and $\prae$ 
(for the respective hierarchies) will not allow r-process 
once collective effects are taken into account and hence will be 
ruled out if one imposes the criteria that $Y_e$ must be less than 
0.5. Therefore, it is expected that the exclusion plot for any 
specific limit on $Y_e$ will change once collective effects are 
taken into account.

\begin{figure}
\includegraphics[width=7cm,height=9cm,angle=270]{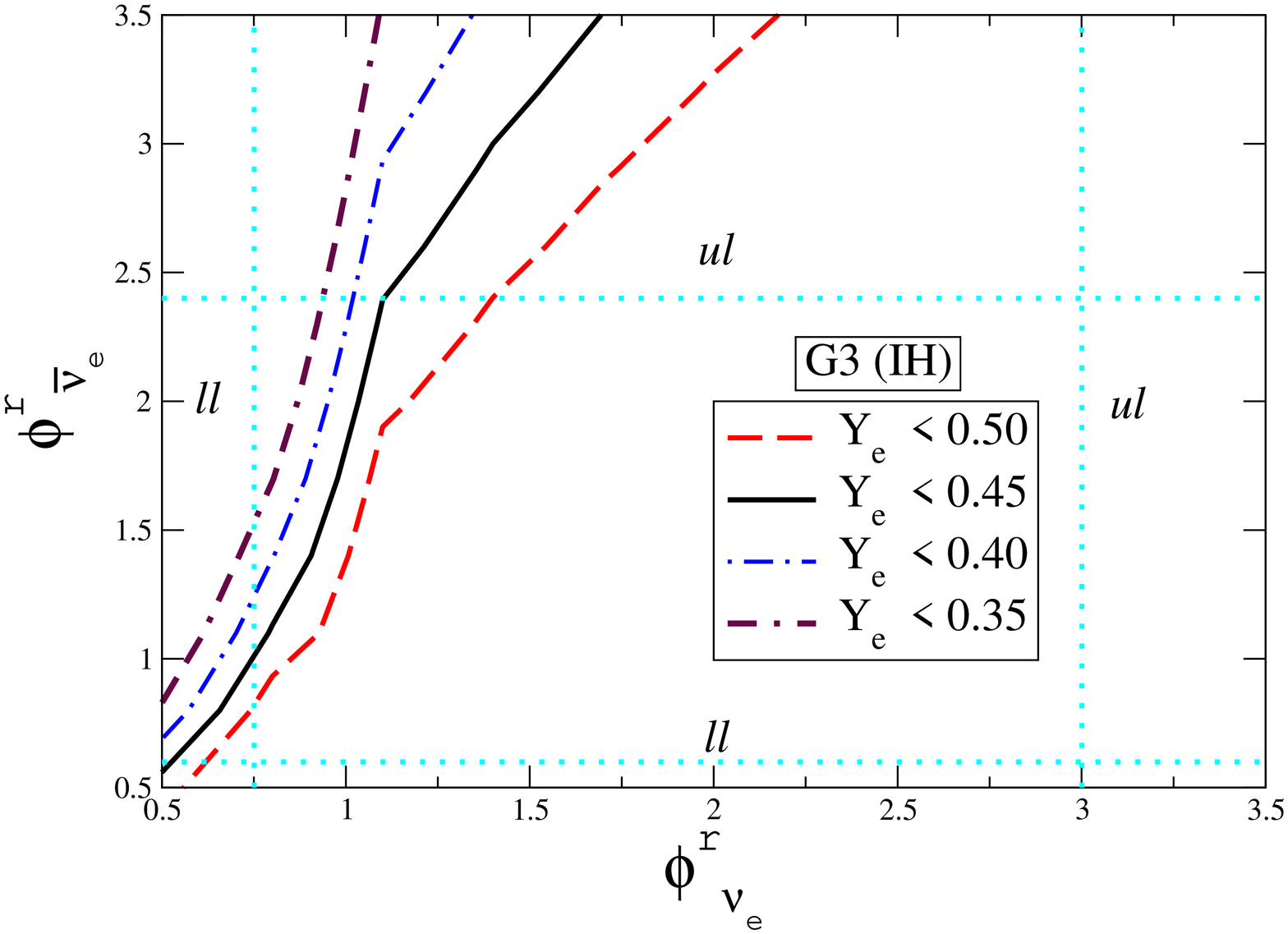}
\hglue -1.0cm
\includegraphics[width=7cm,height=9cm,angle=270]{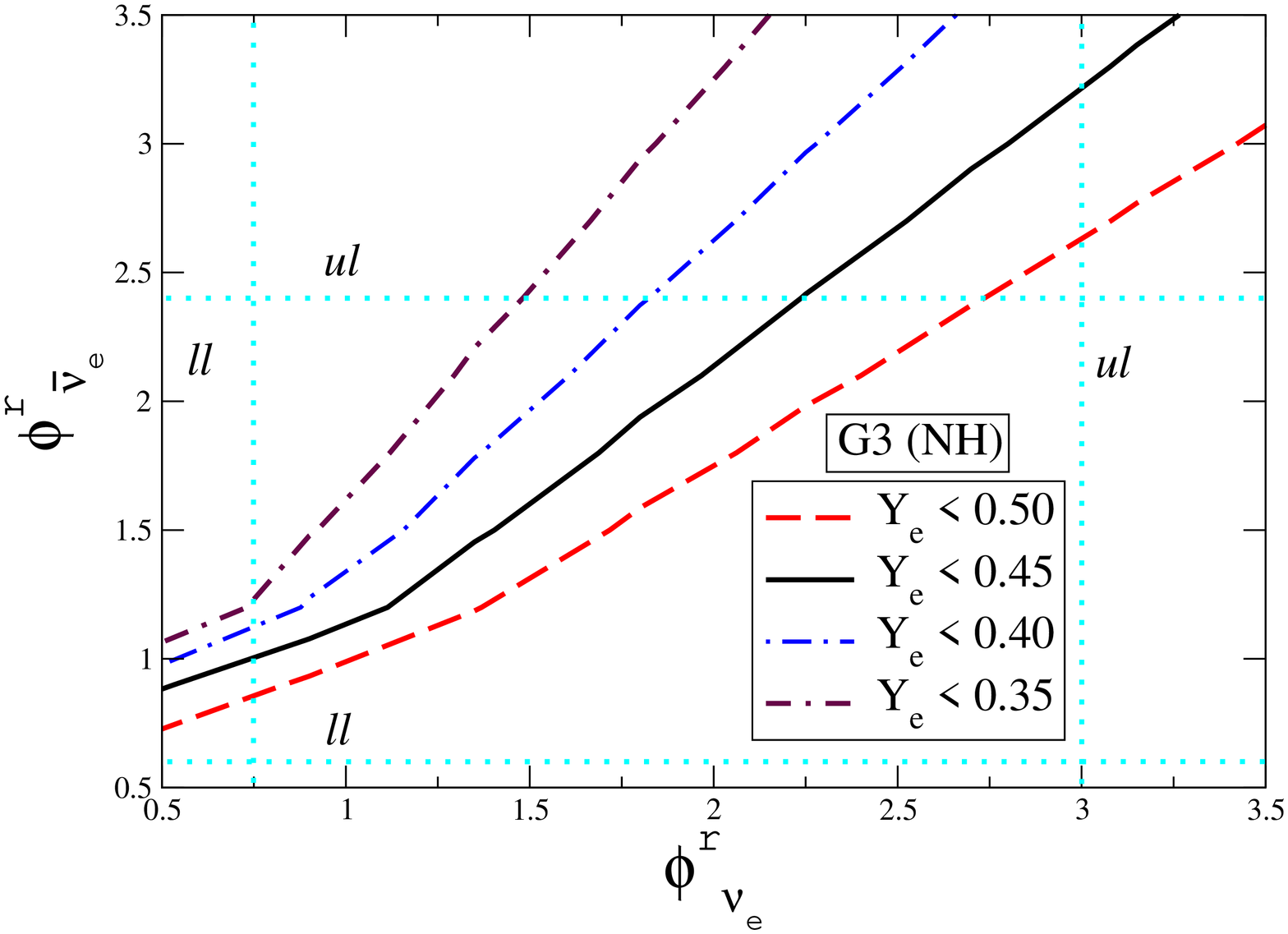}
\caption{\label{fig:flux1}
\footnotesize{
Exclusion plots for NH and IH for model G3. The exclusion 
condition is varied from 
$Y_e < 0.35$ to 0.5. The dotted sky blue lines denote the  
 `$\textit{ll}$' (Lower Limit) and `$\textit{ul}$' (Upper Limit)
 from the assumed two fold uncertainty of $\pre$ and $\prae$,
  for the spectrum model G3. The allowed region is on the 
  left side of the exclusion curves. }
}
\end{figure}

In Figure \ref{Ye<0.45G13LLexcl} we show the 
exclusion plot in the $\pre$-$\prae$ plane for  
$Y_e < 0.45 $. The lines themselves correspond to the case 
when $Y_e=0.45$, while 
the allowed area (which gives $Y_e<0.45$ or $\lambda_{\bar\nu_{e} p}/\lambda_{\nu_{e} n}>1.22 $) 
is to the left of the curves. 
We reiterate that the survival probabilities 
for these plots have been calculated at $r=400$ km. 
We vary $\pre$ and $\prae$ 
in the ranges (0.5,5.0) and  (0.5,3.5)
respectively. We plot exclusion curves for 
both IH and NH, and for all the three spectra models. 
We also plot the exclusion 
curves when collective oscillations are absent 
(WOC) for LL (thick red dashed) and G1/G3 (thick sky blue long dashed). These lines 
correspond to the no conversion case discussed earlier. 
We can see that they are almost straight lines in the 
$\pre$-$\prae$, with $\prae/\pre =1.22* \langle (E+\triangle_{np})^2 \rangle_{\nu_e} / \langle (E-\triangle_{np})^2 \rangle_{\bar\nu_e}$, for the respective spectral model. 
Since the average energy as well as the spectral shape 
(cf. Eq. (\ref{eq:pinched})) for both G1 and G3 are the same, 
the exclusion lines for no oscillation case for them is identical. 
For LL, since the ratio of $\langle (E+\triangle_{np})^2 \rangle_{\nu_e} / \langle (E-\triangle_{np})^2 \rangle_{\anue}$ 
is smaller (0.62) than for G1/G3 (0.88), the exclusion line for no oscillation case 
for LL corresponds to smaller $\prae/\pre$.  
We see from the figure that for NH the effect of 
collective oscillations are mainly in the low 
$\pre$ and $\prae$ region. This conforms to the 
observation in section 2, where we had shown that the 
split energy increased with increasing $\pre$ and/or $\prae$. 
Since the flux begins to fall with increasing energy, 
the impact of collective oscillations fall for higher 
$\pre$ and $\prae$. 
Hence for all the three models, the NH exclusion curves 
at higher relative luminosities agree with the 
WOC ones, as there is no observed 
collective effect there. 
Whereas at lower relative luminosities the curves 
deviate from the WOC ones due to the observed single splits, 
making the allowed region smaller. Again, as seen in Figure 
\ref{Y-evsr}, the effect of collective oscillations is to 
increase $Y_e$ for any given $\prae/\pre$. Since $Y_e$ decreases 
with $\prae/\pre$, the exclusion plots shift to larger $\prae/\pre$ 
values once collective oscillations are switched on. This results 
in the curves shifting left in the $\prae$-$\pre$ plane. 

Figure \ref{Ye<0.45G13LLexcl} shows that for IH, the 
effect of collective oscillations can be very significant 
in constraining the relative luminosities. This can be 
understood from Figures \ref{fig:Split-region} ,\ref{Y-evsr} and Eq \ref{eq:ratio}. In Figure \ref{Y-evsr} we see that the effect of collective oscillation is to shift $Y_e$ to a higher value. Eq. \ref{eq:ratio} shows that this can be balanced by increasing $\phi^{r}_{\bar\nu_e}$ compared to $\phi^{r}_{\nu_e}$. This results in shifting the contour plots to the left of the  $\phi^{r}_{\bar\nu_e}-\phi^{r}_{\nu_e}$ plane in Fig.9.
Since Figures \ref{fig:Split-region} and 
\ref{Ye<0.45G13LLexcl} show the same $\pre$-$\prae$ plane, we can 
see that for IH  once collective oscillations are switched on, 
the only region in this plane which remains allowed is 
the one where we have double splits in the neutrino sector and 
no splits in the antineutrino sector. This is the (II, 0) zone 
where $J_z <0$ and $D_z < 0$.

One understands that stronger constraint on $Y_e$ reduces the allowed parameter space.
Figure 10 shows the exclusion plots for IH and NH for the model G3 for various constraints $Y_e < 0.35$, 0.40, 0.45, 0.50. 
The left panel is for IH and the right one for NH. 
Since higher values of $\pre$ gives higher $Y_e$, as one reduces the 
required value of $Y_e$, $\prae$ gets more constrained. The constraint on 
$\prae$ is relatively weak. But very low values of $\prae$ are not allowed 
as we have seen that lower values of $\prae$ increases the electron fraction.

Some final comments are in order. 
We would like to reiterate here that our analysis and the 
corresponding exclusion plots are meant as a proof of 
principle and merely 
indicate the ranges of 
the allowed fluxes for which
one gets neutron-rich regions for 
r-process in the neutrino driven wind. They show that 
all r-process calculations should take collective effects into account 
and such detailed simulations can be used to extract more rigorous 
bounds on the initial fluxes.

Now we discuss the evolution as a function of time and its
  implication for our results. 
Early times have larger luminosities and may deviate more
from energy equipartition. In realistic simulations in the cooling phase the
luminosity for each species decreases with time as mentioned earlier but it
may be reasonable to assume that the relative luminosities change very slowly
with respect to time. With time the shock moves out, the
neutrinospheres slowly fall in, the matter in the hot
bubble and the wind driven region cools and the constituents change, first
producing alpha-particles and then heavier nuclei. The alpha particles are
strongly bound systems and their excitation by $\nu/\bar\nu$ can be
neglected. The inclusion of $\alpha$-particles in the matter was
considered in \cite{Balantekin:2004ug} and the effect of the nuclear composition involving
heavier nuclei on $Y_{e}$ was looked at in \cite{McLaughlin:1997qi}. A similar study including
spectral splits in a self-consistent manner for the time-evolved system
need to be undertaken separately in future.

One also realizes that there are many uncertainties that exist presently
in the occurrence of the r-process in the supernovae. Firstly large entropy
needed for the development of the r-process, as mentioned earlier, need to
be observed consistently in all one dimensional simulations as well as
in simulations going beyond one dimension. The hot bubble region does have
the problem of having lower entropy \cite{Arnould}. This is compounded by the
inability of simulations to give rise to outgoing shocks with the right
explosion energy. Often the physics understanding comes from the use of
`semi-analytic models' \cite{Arnould} that critically depend on the three quantities
$Y_{e}$, the entropy and the dynamic timescale.

\section{Summary and Conclusions}

Collective flavor oscillations driven by
neutrino-neutrino interaction at the very high density region of core collapse
supernovae control the emitted flux of neutrinos of different flavors.
In the process one or more swaps of flavors for both neutrinos and
antineutrinos take place depending on the initial neutrino flux and 
distributions.
We study the phenomena of spectral splits and consequent 
flavor swaps for different models of 
neutrino spectrum, varying the relative luminosities of neutrinos and 
antineutrinos for both normal and inverted mass hierarchy. 
The effect of spectral splits is found to be 
more pronounced for inverted hierarchy and
depending on the initial luminosities one can get single or dual splits in 
neutrinos and/or antineutrinos. 
For a specific choice of relative luminosity we also find a 
case for inverted hierarchy where the splits are not resolvable numerically 
and is akin to no spectral splits for all practical purposes. 
Single split patterns are also obtained 
for normal hierarchy for some choices of the luminosities. 
Next we consider the impact of the collective oscillations 
and the spectral splits on the electron
fraction $Y_e$,   
which determines if the environment is neutron-rich
and compatible with r-process nucleosynthesis or not. 
The minimal requirement for
r-process is the electron-to-nucleon ratio $Y_{e}< $ 0.5, but a more favorable
condition may be $Y_{e} <$ 0.45/0.40. 
We consider the flavor evolution reduced 
to an effective two flavor model with oscillation between 
$\nu_{e}$ and $\nu_{x}$ and their antiparticles.  The oscillation parameters
are chosen as $\Delta m^{2}=3\times10^{-3}$ eV$^{2}$
and a small effective mixing angle $\theta=10^{-5}$ 
in agreement with
realistic $1-3$ mixing.  
For these values of parameters the ordinary MSW resonances
take place beyond the r-process region and hence 
there will be no effect of neutrino conversions on $Y_e$ due to these. 
However, the inclusion of collective effects can affect the value of 
$Y_e$ even for these values of mass and mixing parameters.

The electron fraction ($Y_{e}$) as a function of the radius of the core is
calculated and it shows an
oscillatory behavior in the bipolar region  
due to collective effects,
before saturating to a constant value which depends on the initial 
luminosities and the pattern of flavor swap. 
Different models of neutrino energy distributions
are used. For each of the distributions initial fluxes of different flavors
are varied and 
constraints on the initial neutrino fluxes consistent 
with successful r-process 
nucleosynthesis are shown in 
exclusion plots for these initial neutrino fluxes. 
While a detailed simulation of the r-process nucleosynthesis 
inside the supernova might bring some changes to the 
exclusion plots, this work illustrates the fact that such 
exclusion plots are possible to achieve.

The variation in the number of spectral splits with the 
variation in the luminosity give rise to different possibilities 
of neutrino and antineutrino spectrum at the detector. 
The constraints on luminosities obtained by ensuring  
r-process nucleosynthesis can provide additional inputs 
in narrowing down  the possible patterns. 

 \textit{Note added}: While this paper was under peer review, a few papers \cite{Friedland:2010, Dasgupta:2010b} have appeared which study the possible effects of a three flavor treatment of the collective oscillations of supernovae neutrinos. The differences of this with a two flavor treatment and their impact on the r-process nucleosynthesis need to be investigated. However from the above studies it is expected that the extra effect due to three flavors may be observable in IH but the changes in NH will be minor.

\section{Acknowledgments} 
S. Chakraborty wishes to thank Basudeb Dasgupta for useful discussions 
and acknowledges hospitality at Physical Research Laboratory and 
Harish-Chandra Research Institute during the development stage of this work. 
S. Choubey and S.G. acknowledge support from the Neutrino Project under the 
XIth plan of Harish-Chandra Research Institute. K.K acknowledges hospitality at 
The Institute of Mathematical Sciences. 
K.K and S. Chakraborty acknowledge support from the
projects `Center for Astroparticle Physics' and `Frontiers of
Theoretical Physics' of Saha Institute of Nuclear Physics.




\end{document}